\documentclass{llncs}
\usepackage[T1]{fontenc}
\usepackage[margin=3.5cm]{geometry}

\usepackage{graphicx}
\usepackage{amsmath}
\usepackage{amsfonts}
\usepackage{multirow}
\usepackage{rotating}
\usepackage{makecell}
\usepackage{booktabs}
\usepackage{colortbl}
\usepackage{hyperref}
\usepackage{listings}
\usepackage{xcolor}
\usepackage{tablefootnote}
\usepackage{amssymb}%
\usepackage{pifont}%
\newcommand{\cmark}{\ding{51}}
\newcommand{\xmark}{\ding{55}}
\usepackage{fontawesome5}
\usepackage{tikz}
\usetikzlibrary{calc,arrows,positioning,shapes,shadows,spy,snakes,plotmarks,matrix,fit,backgrounds}
\usetikzlibrary{external}
\usepackage{pgfplots}
\usepackage{pgfplotstable}
\pgfplotsset{compat=1.4}

\begin{document}
\title{multiGradICON: A Foundation Model \\ for Multimodal Medical Image Registration}
\titlerunning{multiGradICON}

\author{
    Ba\c{s}ar Demir\inst{1}\textsuperscript{(\faEnvelope[regular])} \and
    Lin Tian\inst{1} \and
    Hastings Greer\inst{1} \and
   Roland Kwitt \inst{2} \and \\
    Fran\c{c}ois-Xavier Vialard \inst{3}  \and
    Ra\'ul San Jos\'e Est\'epar \inst{4} \and
    Sylvain Bouix \inst{5} \and \\
    Richard Rushmore \inst{6} \and 
    Ebrahim Ebrahim\inst{7} \and
    Marc Niethammer\inst{1}}

\institute{University of North Carolina at Chapel Hill, USA \and
University of Salzburg, Austria \and
Université Gustave Eiffel, LIGM, France \and
Brigham and Women's Hospital, USA \and
ÉTS Montréal, Canada \and
Boston University, USA \and
Kitware Inc., USA \\
(\faEnvelope[regular])~\email{bdemir@cs.unc.edu}
}
\authorrunning{Demir et al.}

\maketitle

\begin{abstract}
Modern medical image registration approaches predict deformations using deep networks. These approaches achieve state-of-the-art (SOTA) registration accuracy and are generally fast. However, deep learning (DL) approaches %
are, in contrast to conventional non-deep-learning-based approaches, anatomy-specific. Recently, a universal deep registration approach, uniGradICON, has been proposed. However, uniGradICON focuses on monomodal image registration. In this work, we therefore develop multiGradICON as a first step towards universal \emph{multimodal} medical image registration. Specifically, we show that 1) we can train a DL registration model that is suitable for monomodal \emph{and} multimodal registration; 2) loss function randomization can increase multimodal registration accuracy
; and 3) training a model with multimodal data helps multimodal generalization. Our code and the multiGradICON model are available at \href{https://github.com/uncbiag/uniGradICON}{https://github.com/uncbiag/uniGradICON}.

\end{abstract}

\section{Introduction}
Learning-based medical image registration~\cite{chen2023survey,xiao2021review,yang2017quicksilver,balakrishnan2019voxelmorph} has significantly improved over the last decade. Current %
SOTA learning-based deep registration networks~\cite{tian2023gradicon,tian2024unigradicon,mok2020fast,mok2020large} are faster and more accurate than conventional registration methods using numerical optimization. However, learning-based approaches generally 1) focus on monomodal image registration and 2) are trained for a specific anatomical region (often the brain), making them much less generically applicable than conventional approaches. While many monomodal learning-based approaches %
have been developed, uniGradICON~\cite{tian2024unigradicon} is currently the only learning-based approach that is designed to be a \emph{universal} method supporting registrations for different anatomies in \emph{one} model working directly with images and SynthMorph~\cite{hoffmann2021synthmorph} may generalize to other anatomies by training on synthetic shapes. Further, uniGradICON focuses on monomodal registration 
and multimodal generalization is primarily achieved by instance optimization (IO). Hence, our goal is to further close the gap between conventional and learning-based registration approaches by developing multiGradICON, a multimodal generalization of uniGradICON. multiGradICON extends uniGradICON by 1) choosing a multimodal similarity measure%
, 2) incorporating multimodal registration tasks into training, and 3) exploring monomodal, multimodal, and randomized strategies for image similarity loss based on multimodal registration network inputs.

\begin{figure}[t]
    \centering

    \includegraphics[width=0.99\linewidth]{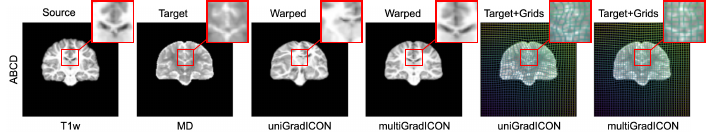}

  \setlength{\belowcaptionskip}{-15pt}
    \caption{Comparison of uni- and multiGradICON on T1w MRI-mean diffusivity~(MD) registration from ABCD. Note the much better matching of the ventricles for multiGradICON.}
    \label{fig:small-figure}
\end{figure}

\noindent Our specific contributions are to show that
\begin{itemize}
\item[1)] while uniGradICON is a strong baseline for monomodal registration, even for unseen modalities, it does not generalize well to multimodal registration when modalities are drastically different (see Fig.~\ref{fig:small-figure} for an example);
\item[2)] training a monomodal model with a squared local normalized cross correlation image similarity loss ($1-\text{LNCC}^2$) does not lead to multimodal generalization in the absence of multimodal registration tasks during training;
\item[3)] multiGradICON allows for multimodal generalization while retaining good monomodal registration accuracy; 
\item[4)] %
similarity loss randomization (i.e., selecting random but identical modalities for image pairs for comparison in the loss)
improves registration accuracy for datasets containing multi-parametric data, even when scalar images are used for inference.
\end{itemize}

\section{Related work}
Following uniGradICON~\cite{tian2024unigradicon} (which builds on ICON~\cite{greer2021icon} and GradICON~\cite{tian2023gradicon}), we focus on non-parametric image registration~\cite{modersitzki2003numerical} where a displacement field is predicted. Such image registrations are generally formulated as a balance between an image (dis)similarity term and a regularizer encouraging spatial smoothness~\cite{modersitzki2003numerical}. By extending uniGradICON, we retain its gradient inverse consistent regularizer (see Sec.~\ref{sec:methodology}) and focus on multimodal alternatives to the negative local normalized cross correlation ($1-\text{LNCC}$) loss used in uniGradICON.

\vskip1ex
\noindent{\textbf{Multimodal similarity measures.}} Multimodal registration requires similarity measures that extend beyond the direct intensity similarities measured by mean squared or absolute error (MSE/MAE) losses. The goal is to assess whether an image pair is spatially aligned even though image appearance may be quite different across modalities. Conventional approaches maximize measures of statistical dependence such as normalized cross correlation (NCC) or squared NCC (if the sign of the correlation is unknown). Even more general statistical dependencies can be captured by maximizing mutual information (MI) between image pairs~\cite{viola1997alignment}. Allowing for more local control, these measures have been extended to local NCC (LNCC; as in uniGradICON) or its squared variant as well as local MI; see~\cite{hermosillo2002variational,modersitzki2003numerical} for an overview of these conventional similarity measures. Such conventional similarity measures have been used to train multimodal deep registration networks~\cite{tian2023gradicon,shen2019networks,guo2019multi}. More modern approaches target similarity measures depending on image self-similarity (such as the modality independent neighborhood descriptor (MIND)~\cite{heinrich2012mind} and its improved version with self-similarity context (MIND-SSC)~\cite{heinrich2013towards}) or local entropy images~\cite{wachinger2012entropy}. The general idea of these more modern similarity measures is to sidestep image differences by using a representation that remains similar even if the underlying image pairs look different. On the extreme end of the similarity measure spectrum, one can then use image segmentations via a Dice loss or try to transform one modality into the other via image synthesis. As such segmentations are generally not directly available and the intensity relationship between image pairs is not a-priori known learning-based approaches are commonly employed.

\vskip1ex

\noindent{\textbf{Learning-based approaches for multimodal similarity quantification.}} Learning better multimodal similarity measures has been explored via non-deep-learning~\cite{lee2009learning} and deep learning approaches~\cite{simonovsky2016deep,yan2018adversarial,cheng2018deep,siebert2022learning,tian2024same,li2023samconvex,mok2024modality}. These multimodal similarity measures can then be used for conventional optimization-based image registration or to train a multimodal deep registration network. In general, multimodal registration can be simplified by 1) converting the multimodal problem into a monomodal one via image synthesis~\cite{xu2020adversarial,roy2013magnetic,qin2019unsupervised}; or 2) if already aligned images of the same subject are available (e.g., multi-parametric sequences in magnetic resonance imaging (MRI)) by only using monomodal pairings for the similarity loss between subjects but multimodal pairings as the input to the learning formulation~\cite{yang2017fast,cao2018deep}; or by 3) training a segmentation model on both modalities and then using the segmentations for the similarity loss~\cite{demirmultimodal}. Several works use the Dice loss computed from image segmentations to train a multimodal registration network~\cite{hu2018weakly,song2022cross,iglesias2024Easyreg,hoffmann2021synthmorph}. Here, segmentations are either obtained manually~\cite{hu2018weakly,song2022cross}, via automatic segmentation algorithms~\cite{iglesias2024Easyreg}, or via label-image pair synthesis~\cite{hoffmann2021synthmorph}. See the recent review articles on deep-learning-based image registration for further details on multimodal formulations~\cite{chen2023survey,xiao2021review}.

\vskip1ex
\noindent\emph{multiGradICON can use any %
multimodal similarity measure. However, as a first step, we focus on $1-\text{LNCC}^2$ as the similarity training loss as it is closely related to uniGradICON's $1-\text{LNCC}$ loss. We use $1-\text{LNCC}^2$ as well as MIND-SSC \cite{heinrich2013towards} for instance optimization. This allows us to keep experiments simple.}

\section{Methodology}
\label{sec:methodology}
We follow the uniGradICON approach with \emph{three key differences}: 1) we adjust the image similarity measure so that it is appropriate for multimodal image registration, 2) we use a larger training dataset that contains multimodal registration tasks, and 3) we explore image similarity loss randomization. We introduce the uniGradICON methodology below and highlight our modifications.

\vskip1ex
\noindent{\textbf{Notation.}} We denote our source and target images by $\left(I^{A}, I^{B} \right)$, and consider them to be functions from an image domain to the real numbers. We denote a registration network by $\Phi_\theta$. This network $\Phi_\theta$ operates on a pair of images to yield a function $\Phi_\theta[I^A, I^B]: \mathbb{R}^N \rightarrow \mathbb{R}^N$ which, when precomposed with the source image, is intended to align the images: $I^A \circ \Phi_\theta[I^A, I^B] \sim I^B\,.$

\subsection{Architecture}
Following the principles outlined in uniGradICON \cite{tian2024unigradicon}, we use the registration network from GradICON \cite{tian2023gradicon}, i.e., we create a multi-step, multi-resolution network using the TwoStep ($\texttt{TS})$ and DownSample ($\texttt{DS}$) operators from \cite{tian2023gradicon}: %
\begin{align}
    \texttt{TS}\{\Psi_\theta^1, \Psi_\theta^2\}[I^A, I^B] &:= \Psi_\theta^1[I^A, I^B] \circ \Psi_\theta^2[I^A \circ \Psi_\theta^1[I^A, I^B], I^B]\,,\\
    \texttt{DS}\{\Psi_\theta\}[I^A, I^B] &:= \Psi_\theta[\texttt{averagePool}(I^A, 2), \texttt{averagePool}(I^B, 2)]\,.
\end{align}
Specifically, we use U-Nets~\cite{cicek2016_3dunet} ($\Psi_\theta^i$) that predict displacement fields and construct the registration network as 
$\Phi_\theta := \texttt{TS}\{\texttt{TS}\{\texttt{DS}\{\texttt{TS}\{\texttt{DS}\{\Psi_\theta^1\}, \Psi_\theta^2\}\}, \Psi_\theta^3\}, \Psi_\theta^4\}\,.$

\subsection{Training losses}
\noindent{\textbf{Baseline.}}
The training loss proposed in GradICON \cite{tian2023gradicon} is 
\begin{equation}
    \begin{aligned}
    \mathcal{L}{=}\mathcal{L}_{\operatorname{sim}}\left(I^{A} \circ \Phi_{\theta}\left[I^{A}, I^{B}\right], I^{B}\right) & + \mathcal{L}_{\operatorname{sim}}\left(I^{B} \circ \Phi_{\theta}\left[I^{B}, I^{A}\right], I^{A}\right) \\
    & +\lambda\left\|\nabla\left(\Phi_{\theta}\left[I^{A}, I^{B}\right] \circ \Phi_{\theta}\left[I^{B}, I^{A}\right]\right)-\mathbf{I}\right\|_F^{2}.
    \end{aligned}
\end{equation}

We use the negative squared localized normalized cross correlation ($1-\text{LNCC}^{2}$) as image similarity measure, $\mathcal{L}_{\text{sim}}(\cdot,\cdot)$, in contrast to uniGradICON's ($1-\text{LNCC}$) loss which assumes locally positive correlations. In contrast, ($1-\text{LNCC}^{2}$) is agnostic to the signs of the correlations. Therefore, it is more appropriate for general multimodal image registration where it is unclear if image intensity pairs are positively or negatively correlated locally.

\vskip1ex
\noindent{\textbf{Image similarity loss randomization}}. Some multimodal medical datasets provide different modalities for the same patient and anatomical region, e.g., for multi-parametric MRI where images from multiple MR sequences are available (this is, for example, the case for brain MRIs of the Human Connectome Project (HCP) or from BratsReg; see Tab.~\ref{table:datasets} for details)\footnote{For notational simplicity, we denote multiple MR sequences also as multimodal rather than multi-sequence data.}. These within-patient images are already aligned since they are derived from the same acquisition. We propose a training strategy to further benefit from these paired multimodal images.

Most deep registration approaches first predict the transformation map to warp the source image to the space of the target image. Image similarity is then calculated between the warped source image and the target image. If the source and target image come from a different modality, this will require a multimodal image similarity measure. If each patient has multiple paired images of different modalities, then the two simplest possible extensions of this approach are to 1) pick one specific modality per patient and proceed with a multimodal image similarity loss or 2) if all patients have the same set of modalities to simply use vector-valued images. The latter approach complicates training a universal model as it would require all datasets to share the same set of modalities which is unrealistic. Instead, we propose extending the former approach. However, we do not pick a specific modality per patient but rather pick a random modality per patient as the input to the deep registration network and another random one to compute the multimodal similarity loss. In expectation, this strategy will train the network with all possible input and loss combinations.

\vskip1ex
 Formally, we assume that our dataset $D=\{P_{i} = \{I_{i}^{1}, I_{i}^{2}, ...,  I_{i}^{m}\}\}_{i \in [n]}$ consists of $m$ scans in different modalities for each patient $P_i$. We first sample a patient pair $\left(P_{A}, P_{B} \right)$. We then uniformly sample a source and target image pair $\left(I^{A},I^{B}\right)$ with $I^{A} \in P_{A}, I^{B} \in P_{B}$ \emph{and} a source and target image pair $\left(I^{A}_L,I^{B}_L\right)$ with $I^{A}_L\in P_{A}, I^{B}_L\in P_{B}$ for similarity loss computations. The pair $\left(I^{A}, I^{B}\right)$ is used for transformation map prediction, and $\left(I_L^{A}, I_L^{B}\right)$ is used for similarity loss calculation. The resulting training loss is
 \begin{equation}
    \begin{aligned}
    \mathcal{L} = \mathcal{L}_{\operatorname{sim}}\left(I_L^{A} \circ \Phi_{\theta}\left[I^{A}, I^{B}\right], I_L^{B}\right)&+  \mathcal{L}_{\operatorname{sim}}\left(I_L^{B} \circ \Phi_{\theta}\left[I^{B}, I^{A}\right], I_L^{A}\right)+ \\
    & +\lambda\left\|\nabla\left(\Phi_{\theta}\left[I^{A}, I^{B}\right] \circ \Phi_{\theta}\left[I^{B}, I^{A}\right]\right)-\mathbf{I}\right\|_F^{2}.
    \end{aligned}
\end{equation}

To explore the effects of choosing the modalities for image similarity calculations we experiment with both sampling $I_L^{A}$ and $I_L^{B}$ randomly or restricting the random sampling to picking the same modality for $I_L^{A}$ and $I_L^{B}$. Similar to our baseline approach, we use ($1-\text{LNCC}^{2}$) as our similarity measure. We train our model with the same hyperparameters as for uniGradICON. We set $\lambda=1.5$.

\subsection{Dataset}
We created a comprehensive training dataset by combining monomodal and multimodal datasets across anatomical regions; see Tab.~\ref{table:datasets} for details. %
We extend the uniGradICON \cite{tian2024unigradicon} training dataset which contains only monomodal datasets. We also used additional modalities available in the uniGradICON datasets. Further, we added new datasets containing a wide range of brain MRI sequences (T1w, T1ce, T2w, FLAIR), contrasts derived from diffusion tensors (fractional anisotropy-FA, mean diffusivity-MD), CT-T1w abdomen MRIs, and DIXON MRIs for fat and water covering anatomical regions across the entire body from neck to knee. Our final corpus is composed of 16 different datasets, contains 5 different anatomical regions (lung, knee, brain, abdomen, pancreas) in addition to a whole body MR dataset, and 12 different image modalities (T1w, T1ce, T2w, T2, FLAIR, DESS, FA, MD, CT, CBCT, Fat/Water DIXON).

\begin{table}
\centering
\caption{Datasets used for training and testing.}
\label{table:datasets}
\resizebox{\linewidth}{!}{%
\begin{tabular}{llccccccc}\toprule
\textbf{Dataset}             & \textbf{Anatom.} & \textbf{\# of}   & \textbf{\# of} & \textbf{Type} & \textbf{Modality}      & \textbf{Label}         & \textbf{\% Training} & \textbf{\% Finetuning }  \\
                             & \textbf{region}  & \textbf{patients} & \textbf{pairs} &               &                        & \textbf{Randomization} &  \textbf{Set}                        &             \textbf{Set}                  \\ 
\hline
COPDGene \cite{regan2011genetic}                   & Lung             & 899                              & 899            & Intra-pat.    & CT                     & \xmark                     & 2.12                     & 8.33                        \\
OAI \cite{nevitt2006osteoarthritis}                        & Knee             & 2532                              & 7,398,400      & Inter-pat.    & DESS/T2 MRI            & \xmark                     & 6.38                     & 12.5                        \\
HCP  \cite{van2012human}                       & Brain            & 1076                              & 4,605,316        & Inter-pat.    & T1w/T2w MRI            & \cmark                    & 6.38                     & 8.33                        \\
L2R-Abdomen \cite{xu2016evaluation}               & Abdomen          & 30                                & 450            & Inter-pat.    & CT                     & \cmark                    & 6.38                     & 6.25                        \\ 
\hline
BratsReg \cite{bratsreg}                  & Brain            &   140                                &       2,240         & Intra-pat.    & T1w/T1ce/T2w/FLAIR MRI & \cmark                    & 21.27                    & 8.33                        \\
ABCD \cite{abcd}                        & Brain            &        302                  &      364,816          & Inter-pat.    & FA/MD                  & \cmark                    & 6.38                     & 0                           \\
L2R-AbdomenMRCT \cite{clark2013cancer,akin2016radiology,CTMRI_KIRP,CTMRI_LIHC}                    & Abdomen          &     97                   &       11,025         & Inter-pat.    & CT/T1w MRI             & \xmark                    & 12.76                    & 6.25                        \\
UK Biobank \cite{ukbiobank}                  & Neck-to-Knee     & 90                             & 194,400          & Inter-pat.    & Fat/Water DIXON        & \cmark                    & 38.29                    & 27.08                       \\ 
\hline
L2R-ThoraxCBCT-train \cite{hugo2016data,hugo2017longitudinal}               & Lung             & 14                                & 1,764             & Inter-pat.    & CT/CBCT                & \xmark                     & 0                        & 8.33                        \\
Pancreatic-CT-CBCT-SEG \cite{hong2021breath} & Pancreas         & 40                               & 720            & Intra-pat.    & CT/CBCT                & \xmark                     & 0                        & 6.25                        \\
ABCD  \cite{abcd}                      & Brain            &        307                          &     1,483,524           & Inter-pat.    & T1w/T2w/FA/MD            & \xmark                     & 0                        & 8.33                        \\ 
\hline
Dirlab-COPDGene \cite{castillo2013reference}             & Lung             & 10                                & 10             & Intra-pat.    & CT                     & -                      & 0                        & 0                           \\
OAI-test \cite{nevitt2006osteoarthritis}                    & Knee             & 301                               & 301            & Inter-pat.    & DESS MRI               & -                      & 0                        & 0                           \\
HCP-test  \cite{van2012human}                    & Brain            & 32                                & 100            & Inter-pat.    & T1w/T2w MRI            & -                      & 0                        & 0                           \\
L2R-NLST-val  \cite{aberle2011reduced,clark2013cancer}               & Lung             & 10                                & 10             & Intra-pat.    & CT                     & -                      & 0                        & 0                           \\
L2R-OASIS-val \cite{marcus2007open,hoopes2021hypermorph}                & Brain            & 20                                & 19             & Inter-pat.    & T1w MRI                & -                      & 0                        & 0                           \\
IXI-test\tablefootnote{\href{https://brain-development.org/ixi-dataset/}{https://brain-development.org/ixi-dataset/}}                     & Brain            & 115                               & 115            & Atlas-pat.    & T1w MRI                & -                      & 0                        & 0                           \\
L2R-ThoraxCBCT-val \cite{hugo2016data,hugo2017longitudinal}                & Lung                              & 3                & 6              & Intra-pat.    & CT/CBCT                & -                      & 0                        & 0                           \\
L2R-AbdomenMRCT-val \cite{clark2013cancer,akin2016radiology,CTMRI_KIRP,CTMRI_LIHC}                & Abdomen                           & 2                & 3              & Intra-pat.    & CT/T1w MRI             & -                      & 0                        & 0                           \\
UK Biobank-test \cite{ukbiobank}             & Neck-to-Knee     & 10                               &        360       &        Inter-pat.       & Fat/Water DIXON        & -                      & 0                        & 0                           \\
Pancreatic-CT-CBCT-SEG \cite{hong2021breath}  & Pancreas         & 40                                & 80            &        Intra-pat.       & CT/CBCT                & -                      & 0                        & 0                          \\ \bottomrule
\end{tabular}
}
\end{table}

\vskip1ex
\noindent{\textbf{Data augmentation.}}
We utilize affine data augmentation which randomly flips the input images and applies random affine transforms, as described in~\cite{tian2023gradicon}. Further, inverting 0-1 normalized CT scans ($1-\text{CT}$) may enhance the segmentation accuracy of CT images when using a network trained on T1w MRIs~\cite{inverse_ct}. This indicates that inverted CT scans may more strongly resemble T1w MRIs. Consequently, we integrate inverted CT scans into our training L2R-Abdomen and L2R-AbdomenMRCT datasets which already include CT images. During finetuning, we only apply random affine augmentation and do not use CT inversion to further fit our model on real image modalities. 

\vskip1ex
\noindent{\textbf{Data balancing.}}
The number of patients, scans, and provided modalities varies across datasets. To ensure a balanced dataset with respect to the number of possible modality-anatomical region combinations, we start by randomly selecting 4,000 image pairs from each dataset. These pairs are then assigned weights based on the dataset they belong to, ensuring an equal representation of observations from each (modality/region) combination during training. We then sample 4,000 3D image pairs per epoch using weighted sampling, consistent with the number of pairs per epoch used in uniGradICON~\cite{tian2024unigradicon}. For \emph{finetuning}, we recompute our weights to account for the additional finetuning datasets;  further, %
we use an anatomic-region-based weighting strategy that balances the number of seen anatomical regions by equally weighting anatomical regions for fine-tuning. Please refer to Tab.~\ref{table:datasets} for the diversity of the datasets and the percentages of each in the training and finetuning sets.

\vskip1ex
\noindent{\textbf{Data preprocessing.}}
We clip the Hounsfield Units (HU) to the range [$-1000$, 1000] for all CT images and then normalize them to [0, 1]. For all MR images (T1w, T1ce, T2w, T2, FLAIR, DESS, DIXON, FA, MD), we clip the maximum intensity at the 99th percentile and then normalize them to [0, 1]. For the pancreatic CT-CBCT dataset, we follow the preprocessing steps outlined in \cite{han2021deep}. We resize all images to a shape of [175, 175, 175] using trilinear interpolation. The spacing across the datasets varies; however, the input pairs (within a dataset) have the same spacing. During inference, we always evaluate our model on the original images by interpolating the transformation maps.

\section{Results}
We conduct experiments to assess multiGradICON's performance and effectiveness compared to the existing monomodal foundational registration model, uniGradICON \cite{tian2024unigradicon}, and optimization-based registration method SyN~\cite{avants2008symmetric}, using default hyperparameters and MI as the similarity measure. We analyze monomodal and multimodal performance separately. 

\vskip1ex
\noindent{\textbf{General hypothesis and questions.}} We hypothesize that multiGradICON can adapt to multimodal data while maintaining comparable monomodal performance to uniGradICON. However, our goal is for multiGradICON to be appropriate for multimodal \emph{and} monomodal registration. Hence, for the monomodal setting, we seek answers to the questions: 1) How does the performance on monomodal datasets in the training set compare between our approach and uniGradICON?; 2) What is the performance difference in additional monomodal datasets that multiGradICON trained on compared to uniGradICON?; 3) Does multiGradICON generalize to unseen cases as well as uniGradICON?

\vskip1ex
\noindent{\textbf{Training design questions.}}
We also investigate the impact of different factors such as training similarity loss selection ($1-\text{LNCC}$ or $1-\text{LNCC}^{2}$), instance optimization similarity loss selection ($1-\text{LNCC}^{2}$ or MIND-SSC), and training loss calculation strategy (baseline or label randomization). For this, we first train uniGradICON with $1-\text{LNCC}^{2}$ %
to investigate the effect of loss selection on generalization to multimodal pairs. Then, we introduce three variants of multiGradICON based on their image similarity loss calculation strategy: 1) the baseline multiGradICON-B approach which uses the same image pairing as input to the network and the loss; 2) multiGradICON-F which uses loss randomization but always samples from the same modality for the loss; 3) multiGradICON-R which also uses loss randomization but allows for sampling from different modalities. Finally, we obtain multiGradICON by further training our best-performing approach multiGradICON-F by including additional datasets to the training set (ThoraxCBCT, Pancreas, ACBD Diffusion (MD or FA)-Structure (T1w or T2w)). For this further training, we do not use $1-\text{CT}$ for data augmentation, we use lung-masked images for the COPDGene dataset, and we recompute the dataset weights (see Tab.~\ref{table:datasets}). We report results for all our methods without instance optimization (w/o IO) or with 50 steps of IO using either $1-\text{LNCC}^{2}$ or MIND-SSC as similarity measures. Note that we perform all of the instance optimization operations on a given pair without any image similarity loss randomization, and we optimize the network parameters for the displacement field using an Adam optimizer with a learning rate of $2\times10^{-5}$.

\subsection{Performance on monomodal registration}
Here, we discuss the performance of multiGradICON on monomodal datasets compared to uniGradICON. We split our evaluations into three categories based on the evaluation datasets: 1) Datasets that exist in both uni- and multiGradICON training sets; 2) Datasets that exist only in the multiGradICON training set; 3) Unseen datasets during training for uniGradICON and multiGradICON.

\begin{table}
\centering
\caption{Performance comparison on the monomodal datasets used for both uni- and multiGradICON training.}
\label{table:uni-monomodal}
\resizebox{\linewidth}{!}{%
\begin{tabular}{clcccccccccc} \toprule
\multicolumn{1}{c}{}    &                   & \multicolumn{4}{c}{Lung}                                              & \multicolumn{2}{c}{Brain}       & \multicolumn{2}{c}{Abdomen}      & \multicolumn{2}{c}{Knee}         \\
\multicolumn{1}{c}{}    &                   & \multicolumn{4}{c}{\cellcolor[RGB]{208,234,208}COPDGene}                                          & \multicolumn{2}{c}{\cellcolor[RGB]{255,228,206}HCP}        & \multicolumn{2}{c}{\cellcolor[RGB]{205,224,238}Abdomen CTCT} & \multicolumn{2}{c}{\cellcolor[RGB]{150,204,198}OAI}          \\
\multicolumn{1}{c}{}    &                   & \multicolumn{2}{c}{CT/CT (masked)} & \multicolumn{2}{c}{CT/CT}       & \multicolumn{2}{c}{T1w/T1w}     & \multicolumn{2}{c}{CT/CT}        & \multicolumn{2}{c}{DESS/DESS}    \\  \cmidrule{3-12}
\multicolumn{1}{c}{}    &                   & mTRE & $\%|J|_{<0}$    & mTRE & $\%|J|_{<0}$ & DICE(\%) & $\%|J|_{<0}$ & DICE(\%) & $\%|J|_{<0}$  & DICE(\%) & $\%|J|_{<0}$  \\ 
\hline
\multirow{7}{*}{w/o IO}                                                                & SyN                           & 8.20          & 0                     & 15.18          & 0                    & 75.8                              & 0                                    & 25.2                                  & 0                                         & 65.7                              & 0                                    \\
                                                                                       & uniGradICON                   & 2.26          & 9.3e-5                & 6.71           & 5.7e-3               & 76.2                              & 6.4e-5                               & 48.3                                  & 3.1e-1                                    & 68.9                              & 6.9e-2                               \\
                                                                                       & uniGradICON-$\text{LNCC}^{2}$ & 2.62          & 9.5e-5                & 6.59           & 1.3e-2               & 76.6                              & 5.9e-5                               & 49.8                                  & 3.2e-1                                    & 69.5                              & 9.0e-2                               \\
                                                                                       & multiGradICON - B             & 5.62          & 1.4e-3                & 6.34           & 3.3e-3               & 75.6                              & 6.4e-5                               & 39.2                                  & 2.2e-1                                    & 64.8                              & 2.1e-2                               \\
                                                                                       & multiGradICON - F             & 5.63          & 2.6e-4                & 6.33           & 1.6e-4               & 75.2                              & 1.3e-5                               & 39.2                                  & 3.7e-2                                    & 65.3                              & 2.0e-2                               \\
                                                                                       & multiGradICON - R             & 5.76          & 4.5e-4                & 6.63           & 1.7e-4               & 74.9                              & 4.8e-6                               & 39.1                                  & 2.3e-2                                    & 65.8                              & 7.7e-3                               \\
                                                                                       & multiGradICON                 & 3.29          & 1.0e-4                & 6.37           & 8.0e-4               & 76.3                              & 1.0e-5                               & 39.4                                  & 2.1e-2                                    & 66.5                              & 5.8e-3                               \\ \hline
\multirow{6}{*}{1-$\text{LNCC}^{2}$} & uniGradICON                   & 1.44          & 2.4e-4                & 2.80           & 1.3e-3               & 78.4                              & 2.0e-4                               & 52.9                                  & 9.4e-1                                    & 69.8                              & 4.8e-2                               \\
                                                                                       & uniGradICON-$\text{LNCC}^{2}$ & 1.46          & 4.2e-4                & 2.97           & 1.7e-3               & 78.7                              & 1.3e-4                               & 53.4                                  & 8.9e-1                                    & 70.2                              & 1.0e-2                               \\
                                                                                       & multiGradICON - B             & 1.75          & 5.4e-5                & 2.65           & 3.6e-4               & 78.1                              & 9.3e-5                               & 46.5                                  & 9.7e-1                                    & 68.4                              & 3.9e-2                               \\
                                                                                       & multiGradICON - F             & 1.73          & 1.1e-5                & 2.64           & 3.7e-5               & 78.1                              & 1.9e-5                               & 48.1                                  & 6.6e-1                                    & 69.3                              & 1.3e-2                               \\
                                                                                       & multiGradICON - R             & 1.76          & 7.4e-6                & 2.67           & 1.4e-4               & 77.8                              & 4.0e-5                               & 47.5                                  & 6.8e-1                                    & 68.9                              & 1.1e-2                               \\
                                                                                       & multiGradICON                 & 1.63          & 3.7e-5                & 2.64           & 1.3e-4               & 78.2                              & 2.8e-5                               & 47.7                                  & 6.4e-1                                    & 69.4                              & 1.1e-2                               \\ \hline
\multirow{6}{*}{MIND-SSC}                                                              & uniGradICON                   & 1.77          & 2.6e-5                & 3.99           & 4.4e-5               & 77.6                              & 3.7e-7                               & 50.8                                  & 4.1e-1                                    & 69.3                              & 4.9e-7                               \\
                                                                                       & uniGradICON-$\text{LNCC}^{2}$ & 1.80          & 6.7e-5                & 4.30           & 1.9e-4               & 77.7                              & 1.6e-6                               & 51.4                                  & 3.8e-1                                    & 69.7                              & 9.8e-5                               \\
                                                                                       & multiGradICON - B             & 2.22          & 0                     & 3.79           & 7.4e-6               & 76.8                              & 0                                    & 42.7                                  & 3.1e-1                                    & 66.6                              & 0                                    \\
                                                                                       & multiGradICON - F             & 2.10          & 0                     & 3.66           & 0                    & 76.9                              & 0                                    & 44.5                                  & 5.9e-3                                    & 67.2                              & 0                                    \\
                                                                                       & multiGradICON - R             & 2.14          & 0                     & 3.75           & 0                    & 76.6                              & 1.8e-7                               & 44.7                                  & 8.5e-3                                    & 67.1                              & 0                                    \\
                                                                                       & multiGradICON                 & 1.95          & 0                     & 3.70           & 0                    & 77.2                              & 0                                    & 44.7                                  & 5.4e-3                                    & 67.5                              & 0                             
             \\ \bottomrule

\end{tabular}
}
\end{table}

\noindent{\textbf{Datasets used for both uni- and multiGradICON training.}}\label{sec:seen-mono} The uni- and multiGradICON training datasets both contain lung (COPDGene CT), brain (HCP T1w MRI), abdomen (L2R Abdomen CT), and knee (OAI MRI) images. Tab.~\ref{table:uni-monomodal} shows that uniGradICON outperforms multiGradICON-B,F,R on the lung, abdomen, and knee datasets based on the initial prediction without instance optimization. This performance difference is around $\sim$3~mm for COPDGene, $\sim$8.5\% Dice score for the abdomen, and $\sim$3\% Dice score for the knee dataset. This result is expected, as uniGradICON is exclusively trained on these datasets and thus has better expertise in these areas. Conversely, multiGradICON is trained on diverse datasets where these specific tasks have lower weight during training. This is further supported by the brain registration results, where multiGradICON-B,F,R show similar performance (within the range of ~0.5 Dice score), with multiGradICON even outperforming uniGradICON by a $\sim$0.1\% Dice score. Since brain datasets are more prevalent in the training sets (e.g., ABCD and BratsReg), multiGradICON performs similarly to uniGradICON on brain registration. After finetuning with anatomical region-based sampling, we observe a performance improvement on the COPDGene dataset, which forms 2.1\% of the training set but is sampled at 8.3\% during finetuning. The performance on the remaining datasets remains similar since their percentages do not change drastically. We observe similar performance improvement on the unseen NLST lung dataset (see Sec.~\ref{sec:unseen-mono}).

\noindent\emph{Instance optimization} with $1-\text{LNCC}^{2}$ narrows the performance gap between uniGradICON and multiGradICON. The difference decreases to $\sim$0.19~mm for COPDGene, $\sim$5\% Dice score for the abdomen, and $\sim$0.4\% Dice score for the knee dataset. Instance optimization with MIND-SSC always under-performs instance optimization with $1-\text{LNCC}^{2}$ across all these datasets.

\noindent\emph{Different multimodal loss strategies} perform similarly for monomodal registration. There are no significant differences in performance, nor is there a clearly dominant method.

\noindent\emph{Lung masking} also affects registration performance. We train our multiGradICON variants using full lung CT images, whereas uniGradICON uses masked images that are zeroed out outside the lung. We observe that a registration model trained without lung masking cannot generalize to register fine details of the lung even if we provide masked lungs during inference. Therefore, during the finetuning process, we further train our model with region of interest (ROI)-masked lung images. After that, we achieve approximately 3~mm improvement in mTRE on masked lung registration without IO. We hypothesize that both the sampling amount in diverse datasets and ROI-masked training are crucial for achieving good registration performance.

\begin{table}[h!]
\centering
\caption{Comparison on monomodal datasets seen by multiGradICON but not by uniGradICON during training.}
\label{table:mm-monomodal}
\resizebox{\linewidth}{!}{%
\begin{tabular}{clcccccccccccccc} \toprule
\multicolumn{1}{c}{}    &                    & \multicolumn{10}{c}{Brain}                                                                                                                                                  & \multicolumn{4}{c}{Neck to Knee}                                           \\
\multicolumn{1}{c}{}    &                    & \multicolumn{2}{c}{\cellcolor[RGB]{205,224,238}HCP}        & \multicolumn{8}{c}{\cellcolor[RGB]{208,234,208}Brats-Reg}                                                                                                            & \multicolumn{4}{c}{\cellcolor[RGB]{255,228,206}UK Biobank}                                             \\
\multicolumn{1}{c}{}    &                    & \multicolumn{2}{c}{T2w/T2w}     & \multicolumn{2}{c}{T1w/T1w}      & \multicolumn{2}{c}{T2w/T2w}      & \multicolumn{2}{c}{T1ce/T1ce}    & \multicolumn{2}{c}{FLAIR/FLAIR} & \multicolumn{2}{c}{WDIXON/WDIXON} & \multicolumn{2}{c}{FDIXON/FDIXON}  \\ \cmidrule{3-16}
\multicolumn{1}{c}{}    &                    & DICE(\%) & $\%|J|_{<0}$ & mTRE & $\%|J|_{<0}$ & mTRE & $\%|J|_{<0}$ & mTRE & $\%|J|_{<0}$ & mTRE & $\%|J|_{<0}$ & DICE(\%) & $\%|J|_{<0}$    & DICE(\%) &$\%|J|_{<0}$     \\ 
\hline
\multirow{7}{*}{w/o IO}              & SyN                           & 75.6                              & 0                                    & 3.50     & 0                & 3.39     & 0                & 3.42      & 0                 & 3.73       & 0                  & 47.7             & 0                    & 43.7            & 0                   \\
                                     & uniGradICON                   & 76.9                              & 5.6e-4                               & 3.27     & 1.0e-3           & 3.31     & 1.2e-3           & 3.24      & 1.4e-3            & 3.83       & 1.9e-3             & 42.2             & 8.1e-3               & 40.0            & 1.6e-2              \\
                                     & uniGradICON-$\text{LNCC}^{2}$ & 77.3                              & 5.0e-4                               & 3.22     & 0                & 3.21     & 0                & 3.13      & 0                 & 3.79       & 0                  & 42.4             & 1.6e-2               & 40.5            & 3.6e-2              \\
                                     & multiGradICON - B             & 76.3                              & 1.0e-4                               & 3.10     & 6.1e-4           & 3.04     & 1.3e-3           & 2.91      & 6.9e-4            & 3.35       & 1.1e-3             & 43.6             & 3.9e-2               & 42.1            & 1.6e-2              \\
                                     & multiGradICON - F             & 76.1                              & 3.3e-6                               & 2.88     & 1.1e-4           & 2.82     & 3.4e-4           & 2.81      & 2.2e-4            & 3.00       & 2.1e-4             & 44.3             & 7.6e-3               & 43.6            & 2.0e-3              \\
                                     & multiGradICON - R             & 75.9                              & 7.4e-6                               & 3.00     & 9.5e-5           & 2.92     & 4.0e-4           & 2.92      & 1.5e-4            & 3.15       & 1.5e-4             & 44.3             & 6.2e-3               & 43.4            & 9.4e-4              \\
                                     & multiGradICON                 & 76.5                              & 7.2e-6                               & 2.90     & 1.2e-4           & 2.86     & 3.6e-4           & 2.82      & 2.0e-4            & 3.05       & 2.0e-4             & 44.8             & 3.1e-3               & 44.1            & 7.6e-4              \\ \hline
\multirow{6}{*}{1-$\text{LNCC}^{2}$} & uniGradICON                   & 77.5                              & 6.1e-4                               & 2.93     & 7.7e-4           & 2.84     & 1.1e-3           & 2.48      & 9.4e-4            & 3.02       & 1.9e-3             & 47.0             & 8.5e-3               & 45.2            & 6.7e-3              \\
                                     & uniGradICON-$\text{LNCC}^{2}$ & 77.9                              & 6.5e-4                               & 2.92     & 8.8e-4           & 2.81     & 1.1e-3           & 2.45      & 9.1e-4            & 2.97       & 1.8e-3             & 46.8             & 1.3e-2               & 45.0            & 1.9e-2              \\
                                     & multiGradICON - B             & 77.2                              & 1.0e-4                               & 2.94     & 7.3e-4           & 2.79     & 1.0e-3           & 2.50      & 7.7e-4            & 2.99       & 1.3e-3             & 47.9             & 5.7e-3               & 46.2            & 5.6e-3              \\
                                     & multiGradICON - F             & 77.3                              & 4.6e-5                               & 2.85     & 5.6e-4           & 2.74     & 1.1e-3           & 2.37      & 5.9e-4            & 2.91       & 1.0e-3             & 48.8             & 3.0e-3               & 48.2            & 3.2e-3              \\
                                     & multiGradICON - R             & 77.1                              & 3.5e-5                               & 2.88     & 7.5e-4           & 2.75     & 1.2e-3           & 2.40      & 9.1e-4            & 2.91       & 1.4e-3             & 48.7             & 2.6e-3               & 48.0            & 2.9e-3              \\
                                     & multiGradICON                 & 77.3                              & 7.7e-5                               & 2.86     & 5.6e-4           & 2.75     & 1.0e-3           & 2.38      & 6.3e-4            & 2.91       & 1.0e-3             & 48.8             & 3.1e-3               & 48.1            & 3.9e-3              \\ \hline
\multirow{6}{*}{MIND-SSC}            & uniGradICON                   & 77.2                              & 7.8e-6                               & 2.70     & 2.1e-6           & 2.54     & 0                & 2.20      & 0                 & 2.62       & 0                  & 45.0             & 0                    & 43.1            & 1.0e-7              \\
                                     & uniGradICON-$\text{LNCC}^{2}$ & 77.5                              & 2.0e-5                               & 2.66     & 3.5e-5           & 2.48     & 0                & 2.14      & 0                 & 2.55       & 2.6e-7             & 44.7             & 5.2e-8               & 43.1            & 0                   \\
                                     & multiGradICON - B             & 76.7                              & 0                                    & 2.72     & 0                & 2.53     & 0                & 2.23      & 9.3e-7            & 2.62       & 0                  & 45.6             & 0                    & 44.1            & 6.7e-7              \\
                                     & multiGradICON - F             & 76.8                              & 0                                    & 2.63     & 0                & 2.49     & 1.3e-7           & 2.18      & 0                 & 2.57       & 0                  & 46.6             & 0                    & 46.2            & 0                   \\
                                     & multiGradICON - R             & 76.6                              & 0                                    & 2.66     & 1.3e-6           & 2.52     & 0                & 2.22      & 1.3e-7            & 2.59       & 6.6e-7             & 46.4             & 0                    & 45.8            & 0                   \\
                                     & multiGradICON                 & 76.9                              & 0                                    & 2.65     & 0                & 2.49     & 0                & 2.19      & 0                 & 2.57       & 2.6e-7             & 46.7             & 0                    & 46.2            & 0    
       \\ \bottomrule
\end{tabular}
}
\end{table}

\noindent{\textbf{Monomodal datasets that only exist in multiGradICON training.}} 
We additionally introduce new monomodal datasets to the multiGradICON training while retaining the existing uniGradICON training datasets. These new monomodal datasets comprise a wide variety of image modalities, such as T2w MRI, FLAIR, and DIXON, which have not been previously seen by uniGradICON. Tab.~\ref{table:mm-monomodal} shows the results across these datasets. Overall, the multiGradICON variants perform slightly better than uniGradICON on the Brats-Reg and UK Biobank datasets, since multiGradICON is trained on these domains. However, both approaches converge to similar performance after 50 steps of IO. These results demonstrate that even on previously unseen monomodal domains, uniGradICON remains a strong baseline, while multiGradICON performs better in initial predictions on the modalities it has seen during training.

\begin{table}[h!]
\centering
\caption{Performance comparison on unseen mono- and multimodal datasets for both uni- and multiGradICON.}
\label{table:unseen}
\resizebox{\linewidth}{!}{%
\begin{tabular}{clcccccccccc}\toprule
\multicolumn{1}{c}{}    &                   & \multicolumn{4}{c}{Lung}                                           & \multicolumn{4}{c}{Brain}                                          & \multicolumn{2}{c}{Pancreas}                \\
\multicolumn{1}{c}{}    &                   & \multicolumn{2}{c}{\cellcolor[RGB]{205,224,238}NLST}        & \multicolumn{2}{c}{\cellcolor[RGB]{208,234,208}ThoraxCBCT}  & \multicolumn{2}{c}{\cellcolor[RGB]{255,228,206}IXI}         & \multicolumn{2}{c}{\cellcolor[RGB]{150,204,198}OASIS}       & \multicolumn{2}{c}{\cellcolor[RGB]{220,204,198}Pancreatic-CT-CBCT-SEG}  \\
\multicolumn{1}{c}{}    &                   & \multicolumn{2}{c}{CT/CT}       & \multicolumn{2}{c}{CT/CBCT}     & \multicolumn{2}{c}{T1w/T1w}     & \multicolumn{2}{c}{T1w/T1w}     & \multicolumn{2}{c}{CT/CBCT}                 \\ \cmidrule{3-12}
\multicolumn{1}{c}{}    &                   & mTRE & $\%|J|_{<0}$ & mTRE & $\%|J|_{<0}$ & DICE(\%) & $\%|J|_{<0}$ & DICE(\%) &$\%|J|_{<0}$ & DICE(\%) & $\%|J|_{<0}$            \\ 
\hline
\multirow{7}{*}{w/o IO}                                                                & SyN                           & 3.04                            & 9.8-e1                                  & 57.4                               & 0                                          & 64.5                              & 1.0e-4                               & 75.6                               & 1.5e-2                                & 78.2                                        & 0                                             \\
                                                                                       & uniGradICON                   & 2.07                            & 4.7e-4                                  & 57.0                               & 4.7e-4                                     & 70.6                              & 7.4e-3                               & 79.0                               & 8.9e-4                                & 81.1                                        & 6.9e-2                                        \\
                                                                                       & uniGradICON-$\text{LNCC}^{2}$ & 2.00                            & 0                                       & 61.0                               & 2.8e-3                                     & 69.7                              & 2.1e-3                               & 79.6                               & 2.8e-3                                & 81.0                                        & 8.1e-2                                        \\
                                                                                       & multiGradICON - B             & 2.74                            & 0                                       & 58.1                               & 3.6e-3                                     & 69.9                              & 2.2e-4                               & 78.6                               & 1.7e-3                                & 80.9                                        & 4.1e-2                                        \\
                                                                                       & multiGradICON - F             & 2.98                            & 0                                       & 57.1                               & 5.0e-3                                     & 70.6                              & 4.7e-5                               & 78.2                               & 8.3e-4                                & 80.3                                        & 9.9e-3                                        \\
                                                                                       & multiGradICON - R             & 2.91                            & 0                                       & 56.8                               & 1.3e-3                                     & 70.3                              & 1.9e-5                               & 77.7                               & 6.4e-4                                & 80.4                                        & 4.7e-3                                        \\
                                                                                       & multiGradICON                 & 2.25                            & 0                                       & 58.0                               & 2.5e-3                                     & 71.8                              & 4.1e-5                               & 78.4                               & 8.4e-4                                & 82.0                                        & 4.7e-3                                        \\ \hline
\multirow{6}{*}{1-$\text{LNCC}^{2}$} & uniGradICON                   & 1.77                            & 8.7e-5                                  & 60.9                               & 2.3e-1                                     & 70.4                              & 1.5e-3                               & 79.7                               & 6.5e-3                                & 82.2                                        & 2.4e-2                                        \\
                                                                                       & uniGradICON-$\text{LNCC}^{2}$ & 1.76                            & 4.8e-5                                  & 62.1                               & 2.5e-2                                     & 70.8                              & 1.6e-3                               & 80.1                               & 9.1e-3                                & 82.0                                        & 4.2e-2                                        \\
                                                                                       & multiGradICON - B             & 1.84                            & 3.1e-4                                  & 60.1                               & 2.0e-2                                     & 70.8                              & 1.0e-3                               & 79.5                               & 5.6e-3                                & 82.0                                        & 1.9e-2                                        \\
                                                                                       & multiGradICON - F             & 1.83                            & 6.1e-4                                  & 60.2                               & 2.7e-1                                     & 70.9                              & 8.0e-4                               & 79.3                               & 5.8e-3                                & 82.3                                        & 2.1e-3                                        \\
                                                                                       & multiGradICON - R             & 1.85                            & 1.7e-5                                  & 60.6                               & 3.1e-1                                     & 71.4                              & 9.4e-4                               & 79.3                               & 6.6e-3                                & 82.3                                        & 1.6e-3                                        \\
                                                                                       & multiGradICON                 & 1.77                            & 2.6e-4                                  & 60.6                               & 2.9e-1                                     & 71.1                              & 1.0e-3                               & 79.4                               & 5.2e-3                                & 82.4                                        & 3.8e-3                                        \\ \hline
\multirow{6}{*}{MIND-SSC}                                                              & uniGradICON                   & 1.87                            & 0                                       & 57.9                               & 2.3e-2                                     & 71.7                              & 1.6e-6                               & 78.9                               & 3.4e-5                                & 82.0                                        & 1.1e-6                                        \\
                                                                                       & uniGradICON-$\text{LNCC}^{2}$ & 1.84                            & 0                                       & 58.3                               & 1.8e-2                                     & 72.2                              & 8.9e-7                               & 79.3                               & 0                                     & 81.8                                        & 1.5e-5                                        \\
                                                                                       & multiGradICON - B             & 1.99                            & 0                                       & 59.4                               & 9.1e-1                                     & 72.0                              & 2.5e-7                               & 78.6                               & 3.1e-6                                & 81.5                                        & 3.4e-6                                        \\
                                                                                       & multiGradICON - F             & 2.01                            & 0                                       & 64.1                               & 0                                          & 72.6                              & 1.2e-7                               & 78.5                               & 0                                     & 81.8                                        & 0                                             \\
                                                                                       & multiGradICON - R             & 2.03                            & 0                                       & 63.0                               & 0                                          & 72.7                              & 0                                    & 78.3                               & 3.1e-6                                & 81.9                                        & 0                                             \\
                                                                                       & multiGradICON                 & 1.93                            & 0                                       & 63.9                               & 0                                          & 72.9                              & 8.9e-7                               & 78.5                               & 0                                     & 82.1                                        & 0 

                      \\ \bottomrule
\end{tabular}
}
\end{table}

\vskip1ex
\noindent{\textbf{Unseen monomodal datasets.}} \label{sec:unseen-mono} We also test on datasets that are never seen during training. Tab.~\ref{table:unseen} shows performance metrics for lung CT-CT NLST and brain T1w MRI registrations from the IXI and OASIS datasets. For NLST, uniGradICON achieves an mTRE of 2.07~mm, whereas the multiGradICON-B,F,R variants show a range between 2.74 and 2.98~mm. We observe a performance improvement on the NLST dataset of approximately 0.73~mm after finetuning. 

\vskip1ex
\noindent\emph{Instance optimization} with $1-\text{LNCC}^{2}$ closes the performance gap on the NLST dataset, with an mTRE of 1.77~mm. However, for brain registration, multiGradICON shows similar performance to uniGradICON, outperforming it by a $\sim$1.2\% Dice score on the IXI dataset and underperforming it by a $\sim$0.6\% Dice score on the OASIS dataset. These results show that multiGradICON scales well to monomodal tasks, providing performance close to that of uniGradICON, which \emph{specializes} in monomodal registration.

\subsection{Performance on multimodal registration}
In this section, we evaluate the multimodal registration performance of multiGradICON on several datasets including a wide variety of anatomical structures and modalities. We again investigate its performance on both seen and unseen datasets. We use uniGradICON as our main comparison model. 

\vskip1ex
\noindent{\textbf{Multimodal training datasets.}} Our training dataset consists of several anatomical regions and modalities such as T1w, T1ce, T2w, FLAIR brain MRIs, CT abdominal scans, and fat and water-weighted DIXON images (see Tab.~\ref{table:datasets} for details). Tab.~ \ref{table:seen-multi} shows registration performances for the HCP, Brats-Reg, Abdomen MR/CT, and UK Biobank datasets. In both the HCP and Brats-Reg datasets, we observe that uniGradICON fails to register pairs that contain images with large appearance differences. For instance, uniGradICON cannot register pairs containing T2w images, even with instance optimization. This is one of the key problems that we aim to solve with multiGradICON. We observe a significant performance improvement on the HCP ($\sim$59\% Dice score improvement), Brats-Reg, and MR-CT ($\sim$9.7\% Dice score improvement) datasets without instance optimization with our multiGradICON approach.

\begin{table}
\centering
\caption{Performance comparison on the multimodal datasets used for multiGradICON training.}
\label{table:seen-multi}
\resizebox{\linewidth}{!}{%
\begin{tabular}{clcccccccccccccccccc}\toprule
\multicolumn{1}{c}{}    &                   & \multicolumn{14}{c}{Brain}                                                                                                                                                                                                                          & \multicolumn{2}{c}{Abdomen}      & \multicolumn{2}{c}{Neck to Knee}     \\
\multicolumn{1}{c}{}    &                   & \multicolumn{2}{c}{\cellcolor[RGB]{205,224,238}HCP}        & \multicolumn{12}{c}{\cellcolor[RGB]{208,234,208}Brats-Reg}                                                                                                                                                                                   & \multicolumn{2}{c}{\cellcolor[RGB]{205,224,238}Abdomen MRCT} & \multicolumn{2}{c}{\cellcolor[RGB]{255,228,206}UK Biobank}       \\
\multicolumn{1}{c}{}    &                   & \multicolumn{2}{c}{T1w/T2w}     & \multicolumn{2}{c}{T1w/T2w}       & \multicolumn{2}{c}{T1w/T1ce}     & \multicolumn{2}{c}{T1w/FLAIR}    & \multicolumn{2}{c}{T2w/T1ce}      & \multicolumn{2}{c}{T2w/FLAIR}    & \multicolumn{2}{c}{T1ce/FLAIR}  & \multicolumn{2}{c}{MR/CT}        & \multicolumn{2}{c}{FDIXON/WDIXON}  \\ \cmidrule{3-20}
\multicolumn{1}{c}{}    &                   & DICE(\%) & $\%|J|_{<0}$ & mTRE  & $\%|J|_{<0}$ & mTRE & $\%|J|_{<0}$ & mTRE & $\%|J|_{<0}$ & mTRE  & $\%|J|_{<0}$ & mTRE & $\%|J|_{<0}$ & mTRE & $\%|J|_{<0}$ & DICE(\%) &$\%|J|_{<0}$  & DICE(\%) & $\%|J|_{<0}$     \\ 
\hline
\multirow{7}{*}{w/o IO}                                          & SyN                           & 71.9                              & 0                                    & 3.74      & 0               & 3.62      & 0                & 3.91      & 0                 & 3.74      & 0                & 3.96      & 0                 & 3.88       & 0                 & 45.0                                  & 0                                         & 33.9                                  & 0                                       \\
                                                                 & uniGradICON                   & 10.0                              & 2.6e-3                               & 13.11     & 3.7e-3          & 4.16      & 2.1e-3           & 6.38      & 2.3e-3            & 12.30     & 4.6e-3           & 8.12      & 5.0e-3            & 6.92       & 3.9e-3            & 50.0                                  & 4.0e-2                                    & 17.5                                  & 2.4e-2                                  \\
                                                                 & uniGradICON-$\text{LNCC}^{2}$ & 14.5                              & 8.6e-4                               & 13.10     & 4.9e-3          & 4.08      & 4.2e-3           & 6.33      & 3.2e-3            & 12.24     & 6.8e-3           & 7.95      & 6.0e-3            & 6.83       & 6.3e-3            & 49.7                                  & 2.6e-2                                    & 18.1                                  & 4.4e-2                                  \\
                                                                 & multiGradICON - B             & 69.2                              & 8.0e-4                               & 3.60      & 1.7e-3          & 3.84      & 8.1e-4           & 4.60      & 1.2e-3            & 3.73      & 2.4e-3           & 4.95      & 2.9e-3            & 4.95       & 2.3e-3            & 58.1                                  & 3.4e-2                                    & 19.7                                  & 1.2e-2                                  \\
                                                                 & multiGradICON - F             & 70.4                              & 5.7e-6                               & 3.04      & 1.5e-4          & 3.05      & 8.8e-5           & 3.46      & 1.3e-4            & 2.93      & 3.0e-4           & 3.49      & 2.9e-4            & 3.53       & 2.0e-4            & 59.4                                  & 9.4e-2                                    & 28.7                                  & 2.6e-3                                  \\
                                                                 & multiGradICON - R             & 70.1                              & 1.6e-5                               & 3.19      & 1.2e-4          & 3.20      & 7.6e-5           & 3.57      & 7.8e-5            & 3.07      & 3.6e-4           & 3.63      & 4.1e-4            & 3.66       & 1.3e-4            & 61.8                                  & 2.2e-2                                    & 27.8                                  & 1.5e-3                                  \\
                                                                 & multiGradICON                 & 70.1                              & 1.5e-5                               & 3.18      & 1.4e-4          & 3.09      & 1.2e-4           & 3.52      & 1.3e-4            & 3.06      & 3.0e-4           & 3.60      & 3.4e-4            & 3.61       & 1.6e-4            & 59.7                                  & 3.3e-2                                    & 28.4                                  & 1.3e-03                                 \\ \hline
\multirow{6}{*}{1-$\text{LNCC}^{2}$} & uniGradICON                   & 7.8                               & 3.3e-3                               & 12.06     & 3.2e-3          & 3.63      & 1.6e-3           & 5.13      & 2.0e-3            & 11.69     & 3.8e-3           & 6.81      & 4.8e-3            & 5.71       & 3.3e-3            & 75.1                                  & 6.9e-1                                    & 17.0                                  & 6.9e-3                                  \\
                                                                 & uniGradICON-$\text{LNCC}^{2}$ & 8.3                               & 2.4e-3                               & 11.94     & 2.7e-3          & 3.59      & 1.8e-3           & 5.05      & 1.9e-3            & 11.48     & 3.2e-3           & 6.60      & 4.5e-3            & 5.63       & 3.2e-3            & 80.3                                  & 5.8e-1                                    & 17.4                                  & 8.6e-3                                  \\
                                                                 & multiGradICON - B             & 72.8                              & 6.3e-4                               & 3.27      & 7.4e-4          & 3.56      & 1.3e-3           & 4.28      & 1.1e-3            & 3.34      & 1.1e-3           & 4.58      & 2.5e-3            & 4.76       & 2.1e-3            & 73.3                                  & 4.4e-1                                    & 18.7                                  & 5.5e-3                                  \\
                                                                 & multiGradICON - F             & 73.2                              & 2.0e-4                               & 3.14      & 8.6e-4          & 3.34      & 1.3e-3           & 3.90      & 9.3e-4            & 2.98      & 9.2e-4           & 4.35      & 1.5e-3            & 4.63       & 1.0e-3            & 75.8                                  & 4.4e-1                                    & 23.2                                  & 2.9e-3                                  \\
                                                                 & multiGradICON - R             & 73.2                              & 2.4e-4                               & 3.15      & 1.0e-3          & 3.33      & 1.8e-3           & 3.87      & 1.4e-3            & 2.99      & 1.3e-3           & 4.36      & 2.2e-3            & 4.49       & 1.6e-3            & 76.6                                  & 3.0e-1                                    & 22.2                                  & 3.0e-3                                  \\
                                                                 & multiGradICON                 & 73.2                              & 1.8e-4                               & 3.17      & 1.0e-3          & 3.35      & 1.4e-3           & 3.89      & 1.0e-3            & 2.99      & 9.8e-4           & 4.40      & 1.8e-3            & 4.60       & 1.2e-3            & 75.5                                  & 4.4e-1                                    & 22.9                                  & 3.5e-3                                  \\ \hline
\multirow{6}{*}{MIND-SSC}                                        & uniGradICON                   & 39.8                              & 1.4e-5                               & 2.66      & 0               & 2.61      & 8.7e-7           & 2.81      & 1.9e-6            & 2.55      & 6.7e-8           & 2.88      & 6.7e-8            & 2.75       & 7.3e-7            & 75.4                                  & 1.9e-4                                    & 20.2                                  & 0                                       \\
                                                                 & uniGradICON-$\text{LNCC}^{2}$ & 53.0                              & 9.5e-6                               & 2.58      & 6.7e-8          & 2.55      & 6.7e-8           & 2.74      & 2.7e-7            & 2.47      & 6.7e-8           & 2.80      & 0                 & 2.68       & 6.7e-8            & 77.4                                  & 1.7e-3                                    & 20.8                                  & 3.1e-7                                  \\
                                                                 & multiGradICON - B             & 74.5                              & 0                                    & 2.70      & 0               & 2.61      & 8.0e-7           & 2.83      & 6.7e-8            & 2.62      & 0                & 2.94      & 0                 & 2.79       & 3.3e-7            & 68.1                                  & 5.0e-3                                    & 23.3                                  & 0                                       \\
                                                                 & multiGradICON - F             & 75.1                              & 0                                    & 2.60      & 0               & 2.48      & 0                & 2.71      & 2.6e-7            & 2.40      & 0                & 2.82      & 0                 & 2.77       & 2.6e-7            & 69.5                                  & 1.4e-3                                    & 40.3                                  & 0                                       \\
                                                                 & multiGradICON - R             & 75.0                              & 0                                    & 2.65      & 0               & 2.51      & 5.3e-7           & 2.76      & 3.9e-7            & 2.42      & 0                & 2.85      & 1.3e-7            & 2.81       & 1.3e-7            & 71.9                                  & 5.6e-4                                    & 39.0                                  & 0                                       \\
                                                                 & multiGradICON                 & 75.1                              & 0                                    & 2.60      & 0               & 2.51      & 1.3e-7           & 2.73      & 3.9e-7            & 2.41      & 0                & 2.82      & 2.6e-7            & 2.79       & 0                 & 70.7                                  & 2.9e-3                                    & 40.1                                  & 0
 \\
\bottomrule                      
\end{tabular}
}
\end{table}

On the other hand, the UK Biobank fat-water weighted DIXON registration is challenging due to the different underlying information across modalities. Although this is a clinically irrelevant scenario since they are acquired together, this result demonstrates the limitation of our approach. We will address this type of registration, where the pairs share the same anatomies but capture entirely different properties, in future work by incorporating semantic information. Additionally, we note that we obtain UK Biobank segmentations using MRSegmentator~\cite{hantze2024mrsegmentator} and evaluate our models directly based on MRSegmentator predictions. Therefore, we can only provide a silver-standard performance metric that may also include possible segmentation errors arising from the MRSegmentator.

We observe that multiGradICON-F and multiGradICON-R outperform multiGradICON-B on the HCP and BraTS-Reg datasets, where loss randomization is applied. This demonstrates that loss randomization improves multimodal registration performance. Specifically, multiGradICON-F performs slightly better than multiGradICON-R across datasets. We hypothesize that selecting the same modality for loss calculations yields better results, because it simplifies the loss computations by only comparing within modalities, whereas comparing between different modalities would require a loss that can effectively work across modalities. 

Additionally, we do not observe any significant multimodal performance improvement in uniGradICON when $1-\text{LNCC}^{2}$ is used as similarity loss during training. This indicates that using appropriate losses for multimodal registration may not lead to multimodal generalization if multimodal datasets were not used during training. We hypothesize that the diversity of the training dataset is more important for generalization.

Moreover, we observe improved registration results with instance optimization (IO) using the MIND-SSC loss in the Brats-Reg dataset compared to IO with $1-\text{LNCC}^{2}$. The Brats-Reg dataset consists of pre-operative and follow-up brain scans that include tumors with varying shapes and resections where each modality captures different tumor properties. As MIND-SSC provides better alignment for these inconsistent structures than $1-\text{LNCC}^{2}$, we conclude that task-specific similarity loss selection can be important for good multimodal registration.

\noindent{\textbf{Unseen multimodal datasets.}} During the training of multiGradICON, we initially did not include the ThoraxCBCT and pancreatic CT-CBCT datasets in our training set. Although the performances of uniGradICON and multiGradICON-B,F,R on these datasets are close to each other, uniGradICON underperforms by $\sim$1\% Dice scores on the ThoraxCBCT dataset and outperforms by $\sim$0.5\% Dice score on the pancreatic CT-CBCT dataset. However, finetuning that incorporates the pancreatic dataset leads to a better Dice score for multiGradICON which outperforms uniGradICON by $\sim$0.9\% Dice score. The performance on the ThoraxCBCT dataset remains similar even when it is included in the training set during finetuning. We hypothesize that our model already converges on CT images during the initial training and that additional CBCT images do not affect the performance since they look sufficiently similar to CT images.

\section{Conclusion}
We developed multiGradICON, a \emph{universal} deep network for \emph{mono- and multimodal} registration. multiGradICON extends uniGradICON, the first deep medical registration network for \emph{monomodal} registration across different anatomies. We observed that uniGradICON remains a strong baseline for monomodal registration but multiGradICON shows improved performance for multimodal registration, in particular, when modalities look so different that not even instance optimization recovers good registrations for uniGradICON. We also demonstrated that similarity loss randomization can bring multimodal registration benefits. Although multiGradICON showed encouraging performance, there are many different avenues for future improvements. For example, we only investigated using $1-\text{LNCC}^{2}$ as a training loss, while many other multimodal similarity measures exist. Further, multiGradICON cannot reliably register between DIXON fat and water images. Although this is somewhat expected, it points to the existence of multimodal datasets that share underlying anatomy but have such large appearance differences that a similarity measure such as $1-\text{LNCC}^{2}$ is insufficient. Using segmentations as part of the image similarity loss may show benefits in such cases. In general, training with segmentations~\cite{xu2019deepatlas} or simulated data for the network input and the similarity loss could be a fruitful avenue for future work. Lastly, just like uniGradICON, multiGradICON is built on top of the exact same network architecture as GradICON. Exploring networks with increased capacity and further increasing training dataset sizes would be desirable. %

\section{Acknowledgements}

This work was supported by NIH grants 1R01AR072013, 1R01AR082684, 1R01EB028283, 1R21MH132982, RF1MH126732, 1R01HL149877, 5R21LM013670, and R01NS125307. The work expresses the views of the authors, not of NIH. Roland Kwitt was supported in part by the Land Salzburg within the EXDIGIT project 20204-WISS/263/6-6022 and projects 0102-F1901166- KZP, 20204-WISS/225/197-2019. Sylvain Bouix was supported in part by Natural Sciences and Engineering Research Council grants RGPIN-2023-05443 and CRC-2022-00183. The knee imaging data were obtained from the controlled access datasets distributed by the Osteoarthritis Initiative (OAI), a data repository housed within the NIMH Data Archive. OAI is a collaborative informatics system created by NIMH and NIAMS to provide a worldwide resource for biomarker identification, scientific investigation, and OA drug development. Dataset identifier: NIMH Data Archive Collection ID: 2343. The brain imaging data were provided by the Human Connectome Project, WU-Minn Consortium (Principal Investigators: David Van Essen and Kamil Ugurbil; 1U54MH091657) funded by the 16 NIH Institutes and Centers that support the NIH Blueprint for Neuroscience Research; and by the McDonnell Center for Systems Neuroscience at Washington University. The lung imaging data were provided by the COPDGene study. Further data was provided by the Learn2Reg challenge, through IXI (Information eXtraction from Images -- EPSRC GR/S21533/02), by  the Brain Tumor Sequence Registration (BraTS-Reg) challenge, and The Cancer Imaging Archive (\url{https://doi.org/10.7937/TCIA.ESHQ-4D90}). Data used in the preparation of this article were also obtained from the Adolescent Brain Cognitive Development$^{\text{SM}}$  (ABCD) Study (\url{https://abcdstudy.org}), held in the NIMH Data Archive (NDA). This is a multisite, longitudinal study designed to recruit more than 10,000 children age 9-10 and follow them over 10 years into early adulthood. The ABCD Study\textsuperscript{\textregistered} is supported by the National Institutes of Health and additional federal partners under award numbers U01DA041048, U01DA050989, U01DA051016, U01DA041022, U01DA051018, U01DA051037, U01DA050987, U01DA041174, U01DA041106, U01DA041117, U01DA041028, U01DA041134, U01DA050988, U01DA051039, U01DA041156, U01DA041025, U01DA041120, U01DA051038, U01DA041148, U01DA041093, U01DA041089, U24DA041123, U24DA041147. A full list of supporters is available at \url{https://abcdstudy.org/federal-partners.html}. A listing of participating sites and a complete listing of the study investigators can be found at \url{https://abcdstudy.org/consortium_members/}. ABCD consortium investigators designed and implemented the study and/or provided data but did not participate in the analysis or writing of this report. This manuscript reflects the views of the authors and may not reflect the opinions or views of the ABCD consortium investigators. The ABCD data used for this work can be found at \url{https://doi.org/10.15154/8v6w-yr62}. This research also used data from the UK Biobank and has been conducted using the UK Biobank Resource under Application Number 22783.

\bibliographystyle{splncs04}
\bibliography{bibliography}

\begin{thebibliography}{10}
\providecommand{\url}[1]{\texttt{#1}}
\providecommand{\urlprefix}{URL }
\providecommand{\doi}[1]{https://doi.org/#1}

\bibitem{aberle2011reduced}
Aberle, D.R., Adams, A.M., Berg, C.D., Black, W.C., Clapp, J.D., Fagerstrom, R.M., Gareen, I.F., Gatsonis, C., Marcus, P.M., Sicks, J., et~al.: Reduced lung-cancer mortality with low-dose computed tomographic screening. The New England journal of medicine  \textbf{365}(5),  395--409 (2011)

\bibitem{akin2016radiology}
Akin, O., Elnajjar, P., Heller, M., Jarosz, R., Erickson, B., Kirk, S., et~al.: Radiology data from the cancer genome atlas kidney renal clear cell carcinoma [{TCGA-KIRC}] collection. The Cancer Imaging Archive  (2016)

\bibitem{avants2008symmetric}
Avants, B.B., Epstein, C.L., Grossman, M., Gee, J.C.: Symmetric diffeomorphic image registration with cross-correlation: evaluating automated labeling of elderly and neurodegenerative brain. MedIA  \textbf{12}(1),  26--41 (2008)

\bibitem{bratsreg}
Baheti, B., Waldmannstetter, D., Chakrabarty, S., Akbari, H., Bilello, M., Wiestler, B., Schwarting, J., Calabrese, E., Rudie, J., Abidi, S., et~al.: The brain tumor sequence registration challenge: establishing correspondence between pre-operative and follow-up {MRI} scans of diffuse glioma patients. arXiv:2112.06979  (2021)

\bibitem{balakrishnan2019voxelmorph}
Balakrishnan, G., Zhao, A., Sabuncu, M.R., Guttag, J., Dalca, A.V.: {VoxelMorph}: a learning framework for deformable medical image registration. TMI  \textbf{38}(8),  1788--1800 (2019)

\bibitem{cao2018deep}
Cao, X., Yang, J., Wang, L., Xue, Z., Wang, Q., Shen, D.: Deep learning based inter-modality image registration supervised by intra-modality similarity. In: MLMI/MICCAI. pp. 55--63 (2018)

\bibitem{abcd}
Casey, B.J., Cannonier, T., Conley, M.I., Cohen, A.O., Barch, D.M., Heitzeg, M.M., Soules, M.E., Teslovich, T., Dellarco, D.V., Garavan, H., et~al.: The adolescent brain cognitive development study: imaging acquisition across 21 sites. Developmental cognitive neuroscience  \textbf{32},  43--54 (2018)

\bibitem{castillo2013reference}
Castillo, R., Castillo, E., Fuentes, D., Ahmad, M., Wood, A.M., et~al.: A reference dataset for deformable image registration spatial accuracy evaluation using the {COPDgene} study archive. Physics in Medicine \& Biology  \textbf{58}(9), ~2861 (2013)

\bibitem{chen2023survey}
Chen, J., Liu, Y., Wei, S., Bian, Z., Subramanian, S., Carass, A., Prince, J.L., Du, Y.: A survey on deep learning in medical image registration: New technologies, uncertainty, evaluation metrics, and beyond. arXiv:2307.15615  (2023)

\bibitem{cheng2018deep}
Cheng, X., Zhang, L., Zheng, Y.: Deep similarity learning for multimodal medical images. Computer Methods in Biomechanics and Biomedical Engineering: Imaging \& Visualization  \textbf{6}(3),  248--252 (2018)

\bibitem{cicek2016_3dunet}
{\c{C}}i{\c{c}}ek, {\"O}., Abdulkadir, A., Lienkamp, S.S., Brox, T., Ronneberger, O.: {3D U-Net}: learning dense volumetric segmentation from sparse annotation. In: MICCAI. pp. 424--432 (2016)

\bibitem{clark2013cancer}
Clark, K., Vendt, B., Smith, K., Freymann, J., Kirby, J., Koppel, P., et~al.: The {Cancer Imaging Archive (TCIA)}: maintaining and operating a public information repository. Journal of digital imaging  \textbf{26},  1045--1057 (2013)

\bibitem{demirmultimodal}
Demir, B., Niethammer, M.: Multimodal image registration guided by few segmentations from one modality. In: MIDL (2024)

\bibitem{CTMRI_LIHC}
Erickson, B.J., Kirk, S., Lee, Y., Bathe, O., Kearns, M., Gerdes, C., et~al.: The cancer genome atlas liver hepatocellular carcinoma collection ({TCGA-LIHC}) (2016)

\bibitem{greer2021icon}
Greer, H., Kwitt, R., Vialard, F.X., Niethammer, M.: {ICON}: Learning regular maps through inverse consistency. In: ICCV (2021)

\bibitem{guo2019multi}
Guo, C.K.: Multi-modal image registration with unsupervised deep learning. Ph.D. thesis, Massachusetts Institute of Technology (2019)

\bibitem{han2021deep}
Han, X., Hong, J., Reyngold, M., Crane, C., Cuaron, J., Hajj, C., Mann, J., Zinovoy, M., Greer, H., Yorke, E., et~al.: Deep-learning-based image registration and automatic segmentation of organs-at-risk in cone-beam ct scans from high-dose radiation treatment of pancreatic cancer. Medical physics  \textbf{48}(6),  3084--3095 (2021)

\bibitem{inverse_ct}
H{\"a}ntze, H., Xu, L., Donle, L., Dorfner, F.J., Hering, A., Adams, L.C., Bressem, K.K.: Improve cross-modality segmentation by treating {MRI} images as inverted {CT} scans. arXiv:2405.03713  (2024)

\bibitem{hantze2024mrsegmentator}
H{\"a}ntze, H., Xu, L., Dorfner, F.J., Donle, L., Truhn, D., Aerts, H., Prokop, M., van Ginneken, B., Hering, A., Adams, L.C., et~al.: Mrsegmentator: Robust multi-modality segmentation of 40 classes in {MRI} and {CT} sequences. arXiv:2405.06463  (2024)

\bibitem{heinrich2012mind}
Heinrich, M.P., Jenkinson, M., Bhushan, M., Matin, T., Gleeson, F.V., Brady, M., Schnabel, J.A.: {MIND}: Modality independent neighbourhood descriptor for multi-modal deformable registration. MedIA  \textbf{16}(7),  1423--1435 (2012)

\bibitem{heinrich2013towards}
Heinrich, M.P., Jenkinson, M., Papie{\.z}, B.W., Brady, S.M., Schnabel, J.A.: Towards realtime multimodal fusion for image-guided interventions using self-similarities. In: MICCAI. pp. 187--194. Springer (2013)

\bibitem{hermosillo2002variational}
Hermosillo, G., Chefd'Hotel, C., Faugeras, O.: Variational methods for multimodal image matching. IJCV  \textbf{50}(3),  329--343 (2002)

\bibitem{hoffmann2021synthmorph}
Hoffmann, M., Billot, B., Greve, D.N., Iglesias, J.E., Fischl, B., Dalca, A.V.: {SynthMorph}: learning contrast-invariant registration without acquired images. TMI  \textbf{41}(3),  543--558 (2021)

\bibitem{hong2021breath}
Hong, J., Reyngold, M., Crane, C., Cuaron, J., Hajj, C., Mann, J., Zinovoy, M., Yorke, E., LoCastro, E., Apte, A., et~al.: Breath-hold {CT} and cone-beam {CT} images with expert manual organ-at-risk segmentations from radiation treatments of locally advanced pancreatic cancer [data set]. TCIA https://doi. org/10.7937/TCIA. ESHQ-4D90  (2021)

\bibitem{hoopes2021hypermorph}
Hoopes, A., Hoffmann, M., Fischl, B., Guttag, J., Dalca, A.V.: Hypermorph: Amortized hyperparameter learning for image registration. In: IPMI (2021)

\bibitem{hu2018weakly}
Hu, Y., Modat, M., Gibson, E., Li, W., Ghavami, N., Bonmati, E., Wang, G., Bandula, S., Moore, C.M., Emberton, M., et~al.: Weakly-supervised convolutional neural networks for multimodal image registration. MedIA  \textbf{49},  1--13 (2018)

\bibitem{hugo2016data}
Hugo, G.D., Weiss, E., Sleeman, W.C., Balik, S., Keall, P.J., Lu, J., Williamson, J.F.: Data from {4D} lung imaging of {NSCLC} patients. The Cancer Imaging Archive  \textbf{10}, ~K9 (2016)

\bibitem{hugo2017longitudinal}
Hugo, G.D., Weiss, E., Sleeman, W.C., Balik, S., Keall, P.J., Lu, J., Williamson, J.F.: A longitudinal four-dimensional computed tomography and cone beam computed tomography dataset for image-guided radiation therapy research in lung cancer. Medical physics  \textbf{44}(2),  762--771 (2017)

\bibitem{iglesias2024Easyreg}
Iglesias, J.E.: A ready-to-use machine learning tool for symmetric multi-modality registration of brain {MRI}. Scientific Reports  \textbf{13}(1), ~6657 (2023)

\bibitem{lee2009learning}
Lee, D., Hofmann, M., Steinke, F., Altun, Y., Cahill, N.D., Scholkopf, B.: Learning similarity measure for multi-modal {3D} image registration. In: CVPR. pp. 186--193 (2009)

\bibitem{li2023samconvex}
Li, Z., Tian, L., Mok, T.C., Bai, X., Wang, P., Ge, J., Zhou, J., Lu, L., Ye, X., Yan, K., et~al.: Samconvex: Fast discrete optimization for {CT} registration using self-supervised anatomical embedding and correlation pyramid. In: MICCAI. pp. 559--569 (2023)

\bibitem{CTMRI_KIRP}
Linehan, M., Gautam, R., Kirk, S., Lee, Y., Roche, C., Bonaccio, E., et~al.: The cancer genome atlas cervical kidney renal papillary cell carcinoma collection ({TCGA-KIRP}) (2016)

\bibitem{marcus2007open}
Marcus, D.S., Wang, T.H., Parker, J., Csernansky, J.G., Morris, J.C., Buckner, R.L.: Open access series of imaging studies ({OASIS}): cross-sectional {MRI} data in young, middle aged, nondemented, and demented older adults. Journal of cognitive neuroscience  \textbf{19}(9),  1498--1507 (2007)

\bibitem{modersitzki2003numerical}
Modersitzki, J.: Numerical methods for image registration (2003)

\bibitem{mok2020fast}
Mok, T.C., Chung, A.: Fast symmetric diffeomorphic image registration with convolutional neural networks. In: CVPR (2020)

\bibitem{mok2020large}
Mok, T.C., Chung, A.C.: Large deformation diffeomorphic image registration with {Laplacian} pyramid networks. In: MICCAI (2020)

\bibitem{mok2024modality}
Mok, T.C., Li, Z., Bai, Y., Zhang, J., Liu, W., Zhou, Y.J., et~al.: Modality-agnostic structural image representation learning for deformable multi-modality medical image registration. arXiv:2402.18933  (2024)

\bibitem{nevitt2006osteoarthritis}
Nevitt, M., Felson, D., Lester, G.: The osteoarthritis initiative. Protocol for the cohort study  \textbf{1}, ~737 (2006)

\bibitem{qin2019unsupervised}
Qin, C., Shi, B., Liao, R., Mansi, T., Rueckert, D., Kamen, A.: Unsupervised deformable registration for multi-modal images via disentangled representations. In: IPMI. pp. 249--261 (2019)

\bibitem{regan2011genetic}
Regan, E.A., Hokanson, J.E., Murphy, J.R., Make, B., Lynch, D.A., Beaty, T.H., et~al.: Genetic epidemiology of {COPD} ({COPDGene}) study design. COPD: Journal of Chronic Obstructive Pulmonary Disease  \textbf{7}(1),  32--43 (2011)

\bibitem{roy2013magnetic}
Roy, S., Carass, A., Prince, J.L.: Magnetic resonance image example-based contrast synthesis. TMI  \textbf{32}(12),  2348--2363 (2013)

\bibitem{shen2019networks}
Shen, Z., Han, X., Xu, Z., Niethammer, M.: Networks for joint affine and non-parametric image registration. In: CVPR (2019)

\bibitem{siebert2022learning}
Siebert, H., Hansen, L., Heinrich, M.P.: Learning a metric for multimodal medical image registration without supervision based on cycle constraints. Sensors  \textbf{22}(3), ~1107 (2022)

\bibitem{simonovsky2016deep}
Simonovsky, M., Guti{\'e}rrez-Becker, B., Mateus, D., Navab, N., Komodakis, N.: A deep metric for multimodal registration. In: MICCAI. pp. 10--18 (2016)

\bibitem{song2022cross}
Song, X., Chao, H., Xu, X., Guo, H., Xu, S., Turkbey, B., Wood, B.J., Sanford, T., Wang, G., Yan, P.: Cross-modal attention for multi-modal image registration. MedIA  \textbf{82},  102612 (2022)

\bibitem{ukbiobank}
Sudlow, C., Gallacher, J., Allen, N., Beral, V., Burton, P., Danesh, J., Downey, P., Elliott, P., Green, J., Landray, M., et~al.: {UK biobank}: an open access resource for identifying the causes of a wide range of complex diseases of middle and old age. PLoS medicine  \textbf{12}(3),  e1001779 (2015)

\bibitem{tian2024unigradicon}
Tian, L., Greer, H., Kwitt, R., Vialard, F.X., Estepar, R.S.J., Bouix, S., Rushmore, R., Niethammer, M.: {uniGradICON}: A foundation model for medical image registration. arXiv:2403.05780  (2024)

\bibitem{tian2023gradicon}
Tian, L., Greer, H., Vialard, F.X., Kwitt, R., Est{\'e}par, R.S.J., Rushmore, R.J., Makris, N., Bouix, S., Niethammer, M.: {GradICON}: Approximate diffeomorphisms via gradient inverse consistency. In: CVPR (2023)

\bibitem{tian2024same}
Tian, L., Li, Z., Liu, F., Bai, X., Ge, J., Lu, L., Niethammer, M., Ye, X., Yan, K., Jin, D.: {SAME++}: A self-supervised anatomical embeddings enhanced medical image registration framework using stable sampling and regularized transformation. arXiv:2311.14986  (2024)

\bibitem{van2012human}
Van~Essen, D.C., Ugurbil, K., Auerbach, E., Barch, D., Behrens, T.E., Bucholz, R., et~al.: The {Human Connectome Project}: a data acquisition perspective. Neuroimage  \textbf{62}(4),  2222--2231 (2012)

\bibitem{viola1997alignment}
Viola, P., Wells~III, W.M.: Alignment by maximization of mutual information. IJCV  \textbf{24}(2),  137--154 (1997)

\bibitem{wachinger2012entropy}
Wachinger, C., Navab, N.: Entropy and {Laplacian} images: Structural representations for multi-modal registration. MedIA  \textbf{16}(1),  1--17 (2012)

\bibitem{xiao2021review}
Xiao, H., Teng, X., Liu, C., Li, T., Ren, G., Yang, R., Shen, D., Cai, J.: A review of deep learning-based three-dimensional medical image registration methods. Quantitative Imaging in Medicine and Surgery  \textbf{11}(12), ~4895 (2021)

\bibitem{xu2020adversarial}
Xu, Z., Luo, J., Yan, J., Pulya, R., Li, X., Wells, W., Jagadeesan, J.: Adversarial uni-and multi-modal stream networks for multimodal image registration. In: MICCAI. pp. 222--232 (2020)

\bibitem{xu2019deepatlas}
Xu, Z., Niethammer, M.: {DeepAtlas}: Joint semi-supervised learning of image registration and segmentation. In: MICCAI. pp. 420--429 (2019)

\bibitem{xu2016evaluation}
Xu, Z., Lee, C.P., Heinrich, M.P., Modat, M., Rueckert, D., Ourselin, S., et~al.: Evaluation of six registration methods for the human abdomen on clinically acquired {CT}. TBE  \textbf{63}(8),  1563--1572 (2016)

\bibitem{yan2018adversarial}
Yan, P., Xu, S., Rastinehad, A.R., Wood, B.J.: Adversarial image registration with application for {MR} and {TRUS} image fusion. In: MLMI/MICCAI. pp. 197--204 (2018)

\bibitem{yang2017fast}
Yang, X., Kwitt, R., Styner, M., Niethammer, M.: Fast predictive multimodal image registration. In: ISBI. pp. 858--862 (2017)

\bibitem{yang2017quicksilver}
Yang, X., Kwitt, R., Styner, M., Niethammer, M.: Quicksilver: Fast predictive image registration--a deep learning approach. NeuroImage  \textbf{158},  378--396 (2017)

\end{thebibliography}

\newpage

\appendix

\vspace*{-50pt}
\section{Additional visualizations}

\begin{figure}[hp]
\vspace*{\fill}
    \centering
    \includegraphics[width=.75\textwidth]{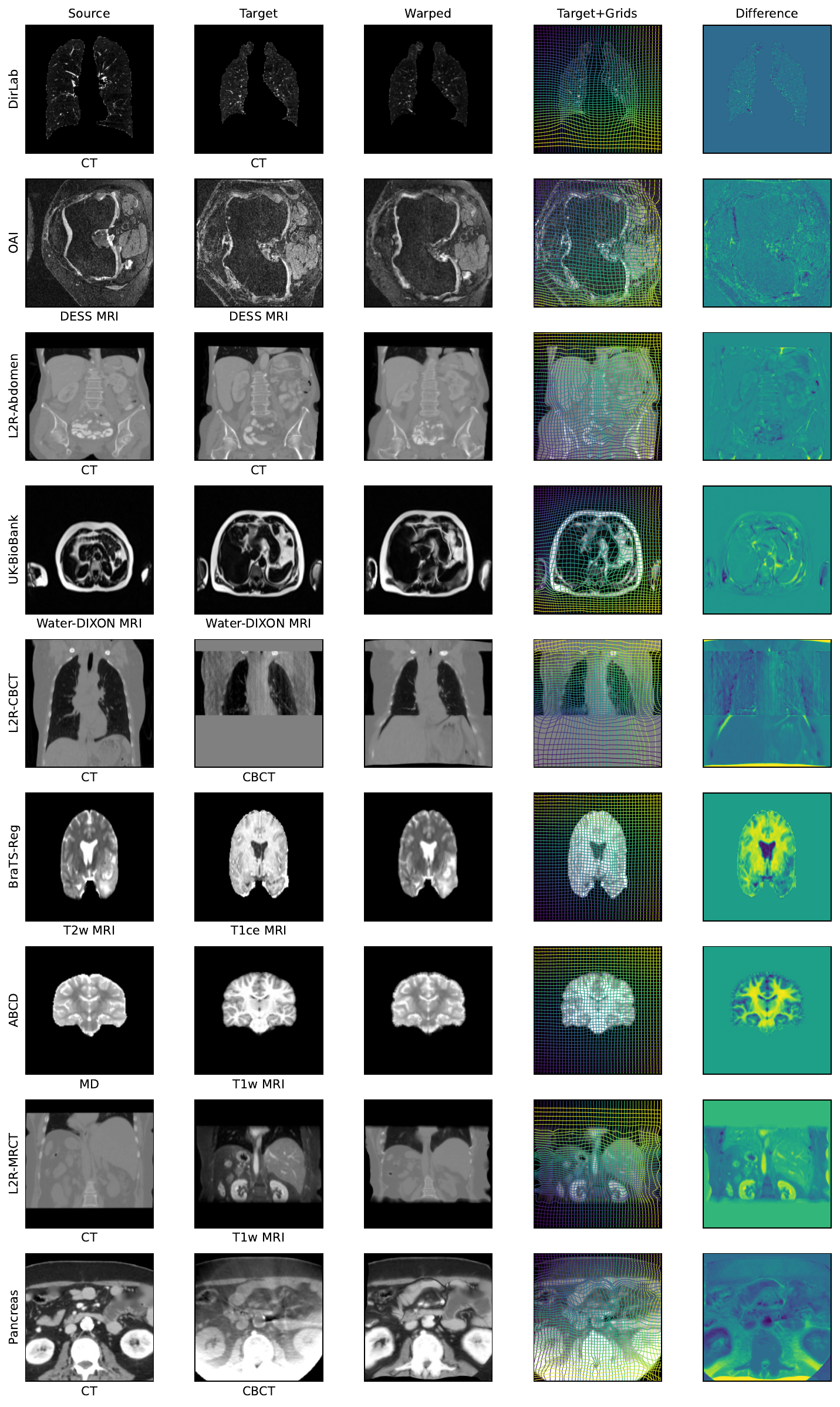}
    \caption{Visualizations of monomodal and multimodal registration results for multiGradICON without instance optimization. The results demonstrate that multiGradICON can handle a wide variety of modalities and anatomies with smooth displacement fields. Note that these images are visualized in their interpolated shapes as they are provided to the network. DIXON images reproduced by kind permission of the UK Biobank\textsuperscript{\textregistered}.}
    \label{fig:main-figure}
\vspace*{\fill}
\end{figure}

Fig.~\ref{fig:main-figure} shows additional registration results for multiGradICON for monomodal and multimodal registration without instance optimization. We observe that multiGradICON can register a wide variety of anatomies and image modalities; with some of the modality/anatomy pairings (e.g., for CT/MRI in the abdomen) being highly challenging.

\section{Erratum}

We identified an error in the multiGradICON training code that affected the results of experiments conducted with image similarity loss randomization. In the intended setup, we sample a source and target image pair, $\left(I^{A},I^{B}\right)$, as input to the model and a separate pair, $\left(I^{A}_L,I^{B}_L\right)$, for similarity loss computations. However, due to the error, the model was trained using the same $\left(I^{A}_L,I^{B}_L\right)$ pair for both model input and similarity loss calculations, instead of sampling two distinct sets of images. Please refer to the Listing~\ref{lst:data-augmenter} for the buggy and the corrected codes.

\begin{lstlisting}[language=Python, basicstyle=\scriptsize, caption={Buggy and corrected implementations of the data augmentation logic. The bug caused the moving image and fixed image variables to be overwritten with moving label and fixed label thereby breaking the independence between the network inputs (moving/fixed image) and the image over which the loss is computed (moving\_label/fixed\_label).}, label={lst:data-augmenter}]
moving_image, fixed_image, moving_image, fixed_image = 
    data_augmenter(moving_image, fixed_image, moving_label, fixed_label)
...             
train_kernel(optimizer, net, moving_image, fixed_image,
    moving_image, fixed_image, writer, iteration)

moving_image, fixed_image, moving_label, fixed_label = 
    data_augmenter(moving_image, fixed_image, moving_label, fixed_label)
...             
train_kernel(optimizer, net, moving_image, fixed_image,
    moving_label, fixed_label, writer, iteration)
\end{lstlisting}

This error caused the multiGradICON - F model to train with input pairs that have identical modalities, since $\left(I^{A}_L,I^{B}_L\right)$ were sampled from the same modality. Consequently, the model performed well on monomodal pairs but failed to generalize to multimodal pairs, which were never seen during training. It further shows that even though the model had seen several different modalities as monomodal pairs during training, it failed to generalize to multimodal pairs composed of those same modalities. For instance, Tables ~\ref{table:mm-monomodal-old} and ~\ref{table:uni-monomodal-old} show that multiGradICON - F is the best model for both the T1w-T1w and T2w-T2w HCP brain registration tasks. However, Table ~\ref{table:seen-multi-old} demonstrates that it fails to generalize to T1w-T2w cases, even though it had seen both T1w and T2w images during training. After fixing the error, multiGradICON - F now performs well on both monomodal and multimodal cases, since it had seen all input pair combinations during training.

This error also affected multiGradICON-R and, by extension, the multiGradICON model training, as image similarity loss randomization was never performed. Instead, the model was trained with multimodal images without loss randomization, same as for multiGradICON-B, since $(I^{A}_L, I^{B}_L)$ were sampled from random modalities and used both as network inputs and for loss randomization. As a result, multiGradICON-B and multiGradICON-R were trained using the same strategy but with different random seeds. 

We retrained all affected models and revised the paper accordingly. In contrast to the previous results, we observed that multiGradICON-F performs slightly better than multiGradICON-R. Therefore, we used multiGradICON-F for multiGradICON training. We observed improvements in multimodal brain registration tasks, where loss randomization plays a more significant role, while the remaining performances remained on par with the previous multiGradICON model. In the interest of academic transparency, we show the previous experimental results below (with affected results highlighted in red) and updated the main manuscript with the now corrected results.
\vspace*{-50pt}

\begin{table}
\centering
\caption{Previous experimental results for Table~\ref{table:uni-monomodal}. An asterisk (*) indicates the affected models.}
\label{table:uni-monomodal-old}
\resizebox{\linewidth}{!}{%
\begin{tabular}{clcccccccccc} \toprule
\multicolumn{1}{c}{}    &                   & \multicolumn{4}{c}{Lung}                                              & \multicolumn{2}{c}{Brain}       & \multicolumn{2}{c}{Abdomen}      & \multicolumn{2}{c}{Knee}         \\
\multicolumn{1}{c}{}    &                   & \multicolumn{4}{c}{\cellcolor[RGB]{208,234,208}COPDGene}                                          & \multicolumn{2}{c}{\cellcolor[RGB]{255,228,206}HCP}        & \multicolumn{2}{c}{\cellcolor[RGB]{205,224,238}Abdomen CTCT} & \multicolumn{2}{c}{\cellcolor[RGB]{150,204,198}OAI}          \\
\multicolumn{1}{c}{}    &                   & \multicolumn{2}{c}{CT/CT (masked)} & \multicolumn{2}{c}{CT/CT}       & \multicolumn{2}{c}{T1w/T1w}     & \multicolumn{2}{c}{CT/CT}        & \multicolumn{2}{c}{DESS/DESS}    \\  \cmidrule{3-12}
\multicolumn{1}{c}{}    &                   & mTRE & $\%|J|_{<0}$    & mTRE & $\%|J|_{<0}$ & DICE(\%) & $\%|J|_{<0}$ & DICE(\%) & $\%|J|_{<0}$  & DICE(\%) & $\%|J|_{<0}$  \\ 
\hline
                                                       & SyN                                        & 8.20                         & 0                              & 15.18                        & 0                               & 75.8                               & 0                                   & 25.2                                   & 0                                        & 65.7                               & 0                                   \\
                                                       & uniGradICON                                & 2.26                         & 9.3e-5                         & 6.71                         & 5.7e-3                          & 76.2                               & 6.4e-5                              & 48.3                                   & 3.1e-1                                   & 68.9                               & 6.9e-2                              \\
                                                       & uniGradICON-$\text{LNCC}^{2}$              & 2.62                         & 9.5e-5                         & 6.59                         & 1.3e-2                          & 76.6                               & 5.9e-5                              & 49.8                                   & 3.2e-1                                   & 69.5                               & 9.0e-2                              \\
                                                       & multiGradICON - B                          & 5.62                         & 1.4e-3                         & 6.34                         & 3.3e-3                          & 75.6                               & 6.4e-5                              & 39.2                                   & 2.2e-1                                   & 64.8                               & 2.1e-2                              \\
                                                       & \cellcolor[HTML]{FD6864}multiGradICON - F* & \cellcolor[HTML]{FD6864}5.12 & \cellcolor[HTML]{FD6864}2.6e-4 & \cellcolor[HTML]{FD6864}5.56 & \cellcolor[HTML]{FD6864}2.0e-3  & \cellcolor[HTML]{FD6864}76.8       & \cellcolor[HTML]{FD6864}3.9e-5      & \cellcolor[HTML]{FD6864}40.5           & \cellcolor[HTML]{FD6864}5.1e-2           & \cellcolor[HTML]{FD6864}66.0       & \cellcolor[HTML]{FD6864}2.0e-2      \\
                                                       & \cellcolor[HTML]{FD6864}multiGradICON - R* & \cellcolor[HTML]{FD6864}6.18 & \cellcolor[HTML]{FD6864}5.0e-4 & \cellcolor[HTML]{FD6864}6.60 & \cellcolor[HTML]{FD6864}3.6e-3  & \cellcolor[HTML]{FD6864}76.4       & \cellcolor[HTML]{FD6864}4.6e-5      & \cellcolor[HTML]{FD6864}40.2           & \cellcolor[HTML]{FD6864}2.7e-1           & \cellcolor[HTML]{FD6864}65.2       & \cellcolor[HTML]{FD6864}5.6e-2      \\
\multirow{-7}{*}{w/o IO}                               & \cellcolor[HTML]{FD6864}multiGradICON*     & \cellcolor[HTML]{FD6864}3.14 & \cellcolor[HTML]{FD6864}7.2e-4 & \cellcolor[HTML]{FD6864}5.85 & \cellcolor[HTML]{FD6864}4.0e-3  & \cellcolor[HTML]{FD6864}76.4       & \cellcolor[HTML]{FD6864}3.7e-5      & \cellcolor[HTML]{FD6864}39.5           & \cellcolor[HTML]{FD6864}8.6e-1           & \cellcolor[HTML]{FD6864}65.4       & \cellcolor[HTML]{FD6864}4.3e-2      \\ \hline
                                                       & uniGradICON                                & 1.44                         & 2.4e-4                         & 2.80                         & 1.3e-3                          & 78.4                               & 2.0e-4                              & 52.9                                   & 9.4e-1                                   & 69.8                               & 4.8e-2                              \\
                                                       & uniGradICON-$\text{LNCC}^{2}$              & 1.46                         & 4.2e-4                         & 2.97                         & 1.7e-3                          & 78.7                               & 1.3e-4                              & 53.4                                   & 8.9e-1                                   & 70.2                               & 1.0e-2                              \\
                                                       & multiGradICON - B                          & 1.75                         & 5.4e-5                         & 2.65                         & 3.6e-4                          & 78.1                               & 9.3e-5                              & 46.5                                   & 9.7e-1                                   & 68.4                               & 3.9e-2                              \\
                                                       & \cellcolor[HTML]{FD6864}multiGradICON - F* & \cellcolor[HTML]{FD6864}1.69 & \cellcolor[HTML]{FD6864}1.4e-4 & \cellcolor[HTML]{FD6864}2.48 & \cellcolor[HTML]{FD6864}5.3e-4  & \cellcolor[HTML]{FD6864}78.4       & \cellcolor[HTML]{FD6864}4.9e-5      & \cellcolor[HTML]{FD6864}48.1           & \cellcolor[HTML]{FD6864}6.2e-1           & \cellcolor[HTML]{FD6864}69.3       & \cellcolor[HTML]{FD6864}1.8e-2      \\
                                                       & \cellcolor[HTML]{FD6864}multiGradICON - R* & \cellcolor[HTML]{FD6864}1.78 & \cellcolor[HTML]{FD6864}5.9e-5 & \cellcolor[HTML]{FD6864}2.91 & \cellcolor[HTML]{FD6864}3.8e-4  & \cellcolor[HTML]{FD6864}78.1       & \cellcolor[HTML]{FD6864}6.7e-5      & \cellcolor[HTML]{FD6864}47.6           & \cellcolor[HTML]{FD6864}7.2e-1           & \cellcolor[HTML]{FD6864}68.4       & \cellcolor[HTML]{FD6864}3.6e-2      \\
\multirow{-6}{*}{1-$\text{LNCC}^{2}$ } & \cellcolor[HTML]{FD6864}multiGradICON*     & \cellcolor[HTML]{FD6864}1.63 & \cellcolor[HTML]{FD6864}1.2e-4 & \cellcolor[HTML]{FD6864}2.92 & \cellcolor[HTML]{FD6864}5.2e-4  & \cellcolor[HTML]{FD6864}78.2       & \cellcolor[HTML]{FD6864}7.6e-5      & \cellcolor[HTML]{FD6864}46.9           & \cellcolor[HTML]{FD6864}6.5e-1           & \cellcolor[HTML]{FD6864}68.2       & \cellcolor[HTML]{FD6864}3.7e-2      \\ \hline
                                                       & uniGradICON                                & 1.77                         & 2.6e-5                         & 3.99                         & 4.4e-5                          & 77.6                               & 3.7e-7                              & 50.8                                   & 4.1e-1                                   & 69.3                               & 4.9e-7                              \\
                                                       & uniGradICON-$\text{LNCC}^{2}$              & 1.80                         & 6.7e-5                         & 4.30                         & 1.9e-4                          & 77.7                               & 1.6e-6                              & 51.4                                   & 3.8e-1                                   & 69.7                               & 9.8e-5                              \\
                                                       & multiGradICON - B                          & 2.22                         & 0                              & 3.79                         & 7.4e-6                          & 76.8                               & 0                                   & 42.7                                   & 3.1e-1                                   & 66.6                               & 0                                   \\
                                                       & \cellcolor[HTML]{FD6864}multiGradICON - F* & \cellcolor[HTML]{FD6864}2.13 & \cellcolor[HTML]{FD6864}1.3e-5 & \cellcolor[HTML]{FD6864}3.39 & \cellcolor[HTML]{FD6864}5.5e-6  & \cellcolor[HTML]{FD6864}77.4       & \cellcolor[HTML]{FD6864}1.8e-7      & \cellcolor[HTML]{FD6864}44.5           & \cellcolor[HTML]{FD6864}8.6e-3           & \cellcolor[HTML]{FD6864}67.4       & \cellcolor[HTML]{FD6864}0           \\
                                                       & \cellcolor[HTML]{FD6864}multiGradICON - R* & \cellcolor[HTML]{FD6864}2.25 & \cellcolor[HTML]{FD6864}0      & \cellcolor[HTML]{FD6864}3.92 & \cellcolor[HTML]{FD6864}7.4e-6  & \cellcolor[HTML]{FD6864}76.9       & \cellcolor[HTML]{FD6864}1.8e-7      & \cellcolor[HTML]{FD6864}44.4           & \cellcolor[HTML]{FD6864}3.6e-2           & \cellcolor[HTML]{FD6864}66.4       & \cellcolor[HTML]{FD6864}0           \\
\multirow{-6}{*}{MIND-SSC}                             & \cellcolor[HTML]{FD6864}multiGradICON*     & \cellcolor[HTML]{FD6864}2.03 & \cellcolor[HTML]{FD6864}1.3e-5 & \cellcolor[HTML]{FD6864}3.99 & \cellcolor[HTML]{FD6864}1.86e-6 & \cellcolor[HTML]{FD6864}77.1       & \cellcolor[HTML]{FD6864}7.4e-7      & \cellcolor[HTML]{FD6864}43.5           & \cellcolor[HTML]{FD6864}3.0e-2           & \cellcolor[HTML]{FD6864}66.4       & \cellcolor[HTML]{FD6864}0           \\ \hline

\end{tabular}
}
\end{table}

\begin{table}[h!]
\centering
\caption{Previous experimental results for Table~\ref{table:mm-monomodal}. An asterisk (*) indicates the affected models.}
\label{table:mm-monomodal-old}
\resizebox{\linewidth}{!}{%
\begin{tabular}{clcccccccccccccc} \toprule
\multicolumn{1}{c}{}    &                    & \multicolumn{10}{c}{Brain}                                                                                                                                                  & \multicolumn{4}{c}{Neck to Knee}                                           \\
\multicolumn{1}{c}{}    &                    & \multicolumn{2}{c}{\cellcolor[RGB]{205,224,238}HCP}        & \multicolumn{8}{c}{\cellcolor[RGB]{208,234,208}Brats-Reg}                                                                                                            & \multicolumn{4}{c}{\cellcolor[RGB]{255,228,206}UK Biobank}                                             \\
\multicolumn{1}{c}{}    &                    & \multicolumn{2}{c}{T2w/T2w}     & \multicolumn{2}{c}{T1w/T1w}      & \multicolumn{2}{c}{T2w/T2w}      & \multicolumn{2}{c}{T1ce/T1ce}    & \multicolumn{2}{c}{FLAIR/FLAIR} & \multicolumn{2}{c}{WDIXON/WDIXON} & \multicolumn{2}{c}{FDIXON/FDIXON}  \\ \cmidrule{3-16}
\multicolumn{1}{c}{}    &                    & DICE(\%) & $\%|J|_{<0}$ & mTRE & $\%|J|_{<0}$ & mTRE & $\%|J|_{<0}$ & mTRE & $\%|J|_{<0}$ & mTRE & $\%|J|_{<0}$ & DICE(\%) & $\%|J|_{<0}$    & DICE(\%) &$\%|J|_{<0}$     \\ 
\hline
                        & SyN               & 75.6 & 0                                                    & 3.50    & 0                    & 3.39    & 0                    & 3.42    & 0                      & 3.73    & 0                        & 47.7 & 0                          & 43.7 & 0                           \\
                                                 & SyN                                        & 75.6                               & 0                                   & 3.50                         & 0                              & 3.39                         & 0                              & 3.42                         & 0                              & 3.73                         & 0                              & 47.7                         & 0                              & 43.7                         & 0                              \\
                                                 & uniGradICON                                & 76.9                               & 5.6e-4                              & 3.27                         & 1.0e-3                         & 3.31                         & 1.2e-3                         & 3.24                         & 1.4e-3                         & 3.83                         & 1.9e-3                         & 42.2                         & 8.1e-3                         & 40.0                         & 1.6e-2                         \\
                                                 & uniGradICON-$\text{LNCC}^{2}$              & 77.3                               & 5.0e-4                              & 3.22                         & 0                              & 3.21                         & 0                              & 3.13                         & 0                              & 3.79                         & 0                              & 42.4                         & 1.6e-2                         & 40.5                         & 3.6e-2                         \\
                                                 & multiGradICON - B                          & 76.3                               & 1.0e-4                              & 3.10                         & 6.1e-4                         & 3.04                         & 1.3e-3                         & 2.91                         & 6.9e-4                         & 3.35                         & 1.1e-3                         & 43.6                         & 3.9e-2                         & 42.1                         & 1.6e-2                         \\
                                                 & \cellcolor[HTML]{FD6864}multiGradICON - F* & \cellcolor[HTML]{FD6864}77.2       & \cellcolor[HTML]{FD6864}6.8e-5      & \cellcolor[HTML]{FD6864}2.94 & \cellcolor[HTML]{FD6864}7.4e-4 & \cellcolor[HTML]{FD6864}2.95 & \cellcolor[HTML]{FD6864}1.6e-3 & \cellcolor[HTML]{FD6864}2.73 & \cellcolor[HTML]{FD6864}9.2e-4 & \cellcolor[HTML]{FD6864}3.14 & \cellcolor[HTML]{FD6864}1.3e-3 & \cellcolor[HTML]{FD6864}45.5 & \cellcolor[HTML]{FD6864}1.8e-2 & \cellcolor[HTML]{FD6864}44.5 & \cellcolor[HTML]{FD6864}1.2e-2 \\
                                                 & \cellcolor[HTML]{FD6864}multiGradICON - R* & \cellcolor[HTML]{FD6864}76.6       & \cellcolor[HTML]{FD6864}6.5e-5      & \cellcolor[HTML]{FD6864}3.07 & \cellcolor[HTML]{FD6864}1.2e-4 & \cellcolor[HTML]{FD6864}3.04 & \cellcolor[HTML]{FD6864}2.1e-3 & \cellcolor[HTML]{FD6864}2.87 & \cellcolor[HTML]{FD6864}1.3e-3 & \cellcolor[HTML]{FD6864}3.33 & \cellcolor[HTML]{FD6864}1.9e-3 & \cellcolor[HTML]{FD6864}43.6 & \cellcolor[HTML]{FD6864}4.4e-2 & \cellcolor[HTML]{FD6864}41.9 & \cellcolor[HTML]{FD6864}3.7e-2 \\
\multirow{-6}{*}{w/o IO}                         & \cellcolor[HTML]{FD6864}multiGradICON*     & \cellcolor[HTML]{FD6864}76.5       & \cellcolor[HTML]{FD6864}1.4e-4      & \cellcolor[HTML]{FD6864}3.06 & \cellcolor[HTML]{FD6864}8.8e-4 & \cellcolor[HTML]{FD6864}3.04 & \cellcolor[HTML]{FD6864}1.6e-3 & \cellcolor[HTML]{FD6864}2.90 & \cellcolor[HTML]{FD6864}1.2e-3 & \cellcolor[HTML]{FD6864}3.38 & \cellcolor[HTML]{FD6864}1.6e-3 & \cellcolor[HTML]{FD6864}43.8 & \cellcolor[HTML]{FD6864}2.3e-2 & \cellcolor[HTML]{FD6864}42.1 & \cellcolor[HTML]{FD6864}2.0e-2 \\ \hline
                                                 & uniGradICON                                & 77.5                               & 6.1e-4                              & 2.93                         & 7.7e-4                         & 2.84                         & 1.1e-3                         & 2.48                         & 9.4e-4                         & 3.02                         & 1.9e-3                         & 47.0                         & 8.5e-3                         & 45.2                         & 6.7e-3                         \\
                                                 & uniGradICON-$\text{LNCC}^{2}$              & 77.9                               & 6.5e-4                              & 2.92                         & 8.8e-4                         & 2.81                         & 1.1e-3                         & 2.45                         & 9.1e-4                         & 2.97                         & 1.8e-3                         & 46.8                         & 1.3e-2                         & 45.0                         & 1.9e-2                         \\
                                                 & multiGradICON - B                          & 77.2                               & 1.0e-4                              & 2.94                         & 7.3e-4                         & 2.79                         & 1.0e-3                         & 2.50                         & 7.7e-4                         & 2.99                         & 1.3e-3                         & 47.9                         & 5.7e-3                         & 46.2                         & 5.6e-3                         \\
                                                 & \cellcolor[HTML]{FD6864}multiGradICON - F* & \cellcolor[HTML]{FD6864}77.6       & \cellcolor[HTML]{FD6864}1.3e-4      & \cellcolor[HTML]{FD6864}2.92 & \cellcolor[HTML]{FD6864}8.1e-4 & \cellcolor[HTML]{FD6864}2.80 & \cellcolor[HTML]{FD6864}1.2e-3 & \cellcolor[HTML]{FD6864}2.49 & \cellcolor[HTML]{FD6864}9.9e-4 & \cellcolor[HTML]{FD6864}2.95 & \cellcolor[HTML]{FD6864}1.6e-3 & \cellcolor[HTML]{FD6864}48.6 & \cellcolor[HTML]{FD6864}5.2e-3 & \cellcolor[HTML]{FD6864}47.6 & \cellcolor[HTML]{FD6864}6.7e-3 \\
                                                 & \cellcolor[HTML]{FD6864}multiGradICON - R* & \cellcolor[HTML]{FD6864}77.2       & \cellcolor[HTML]{FD6864}1.6e-4      & \cellcolor[HTML]{FD6864}2.92 & \cellcolor[HTML]{FD6864}8.8e-4 & \cellcolor[HTML]{FD6864}2.79 & \cellcolor[HTML]{FD6864}1.1e-3 & \cellcolor[HTML]{FD6864}2.47 & \cellcolor[HTML]{FD6864}9.7e-4 & \cellcolor[HTML]{FD6864}2.99 & \cellcolor[HTML]{FD6864}1.6e-3 & \cellcolor[HTML]{FD6864}47.1 & \cellcolor[HTML]{FD6864}6.0e-3 & \cellcolor[HTML]{FD6864}45.4 & \cellcolor[HTML]{FD6864}8.1e-3 \\
\multirow{-6}{*}{1-$\text{LNCC}^{2}$} & \cellcolor[HTML]{FD6864}multiGradICON*     & \cellcolor[HTML]{FD6864}77.2       & \cellcolor[HTML]{FD6864}2.6e-4      & \cellcolor[HTML]{FD6864}2.92 & \cellcolor[HTML]{FD6864}9.7e-4 & \cellcolor[HTML]{FD6864}2.81 & \cellcolor[HTML]{FD6864}1.2e-3 & \cellcolor[HTML]{FD6864}2.48 & \cellcolor[HTML]{FD6864}1.0e-3 & \cellcolor[HTML]{FD6864}2.99 & \cellcolor[HTML]{FD6864}1.9e-3 & \cellcolor[HTML]{FD6864}47.3 & \cellcolor[HTML]{FD6864}6.9e-3 & \cellcolor[HTML]{FD6864}45.8 & \cellcolor[HTML]{FD6864}1.0e-2 \\ \hline
                                                 & uniGradICON                                & 77.2                               & 7.8e-6                              & 2.70                         & 2.1e-6                         & 2.54                         & 0                              & 2.20                         & 0                              & 2.62                         & 0                              & 45.0                         & 0                              & 43.1                         & 1.0e-7                         \\
                                                 & uniGradICON-$\text{LNCC}^{2}$              & 77.5                               & 2.0e-5                              & 2.66                         & 3.5e-5                         & 2.48                         & 0                              & 2.14                         & 0                              & 2.55                         & 2.6e-7                         & 44.7                         & 5.2e-8                         & 43.1                         & 0                              \\
                                                 & multiGradICON - B                          & 76.7                               & 0                                   & 2.72                         & 0                              & 2.53                         & 0                              & 2.23                         & 9.3e-7                         & 2.62                         & 0                              & 45.6                         & 0                              & 44.1                         & 6.7e-7                         \\
                                                 & \cellcolor[HTML]{FD6864}multiGradICON - F* & \cellcolor[HTML]{FD6864}77.2       & \cellcolor[HTML]{FD6864}7.4e-7      & \cellcolor[HTML]{FD6864}2.72 & \cellcolor[HTML]{FD6864}0      & \cellcolor[HTML]{FD6864}2.52 & \cellcolor[HTML]{FD6864}0      & \cellcolor[HTML]{FD6864}2.23 & \cellcolor[HTML]{FD6864}0      & \cellcolor[HTML]{FD6864}2.63 & \cellcolor[HTML]{FD6864}3.9e-7 & \cellcolor[HTML]{FD6864}46.8 & \cellcolor[HTML]{FD6864}0      & \cellcolor[HTML]{FD6864}46.0 & \cellcolor[HTML]{FD6864}0      \\
                                                 & \cellcolor[HTML]{FD6864}multiGradICON - R* & \cellcolor[HTML]{FD6864}76.6       & \cellcolor[HTML]{FD6864}0           & \cellcolor[HTML]{FD6864}2.69 & \cellcolor[HTML]{FD6864}0      & \cellcolor[HTML]{FD6864}2.53 & \cellcolor[HTML]{FD6864}1.3e-7 & \cellcolor[HTML]{FD6864}2.23 & \cellcolor[HTML]{FD6864}1.3e-7 & \cellcolor[HTML]{FD6864}2.64 & \cellcolor[HTML]{FD6864}6.2e-6 & \cellcolor[HTML]{FD6864}44.8 & \cellcolor[HTML]{FD6864}2.1e-7 & \cellcolor[HTML]{FD6864}43.5 & \cellcolor[HTML]{FD6864}0      \\
\multirow{-6}{*}{MIND-SSC}                       & \cellcolor[HTML]{FD6864}multiGradICON*     & \cellcolor[HTML]{FD6864}76.6       & \cellcolor[HTML]{FD6864}7.4e-7      & \cellcolor[HTML]{FD6864}2.69 & \cellcolor[HTML]{FD6864}0      & \cellcolor[HTML]{FD6864}2.53 & \cellcolor[HTML]{FD6864}0      & \cellcolor[HTML]{FD6864}2.22 & \cellcolor[HTML]{FD6864}2.6e-7 & \cellcolor[HTML]{FD6864}2.62 & \cellcolor[HTML]{FD6864}1.3e-7 & \cellcolor[HTML]{FD6864}45.1 & \cellcolor[HTML]{FD6864}1.6e-7 & \cellcolor[HTML]{FD6864}43.7 & \cellcolor[HTML]{FD6864}0      \\ \hline
\end{tabular}%
}
\end{table}

\begin{table}[h!]
\centering
\caption{Previous experimental results for Table~\ref{table:unseen}. An asterisk (*) indicates the affected models.}
\label{table:unseen-old}
\resizebox{\linewidth}{!}{%
\begin{tabular}{clcccccccccc}\toprule
\multicolumn{1}{c}{}    &                   & \multicolumn{4}{c}{Lung}                                           & \multicolumn{4}{c}{Brain}                                          & \multicolumn{2}{c}{Pancreas}                \\
\multicolumn{1}{c}{}    &                   & \multicolumn{2}{c}{\cellcolor[RGB]{205,224,238}NLST}        & \multicolumn{2}{c}{\cellcolor[RGB]{208,234,208}ThoraxCBCT}  & \multicolumn{2}{c}{\cellcolor[RGB]{255,228,206}IXI}         & \multicolumn{2}{c}{\cellcolor[RGB]{150,204,198}OASIS}       & \multicolumn{2}{c}{\cellcolor[RGB]{220,204,198}Pancreatic-CT-CBCT-SEG}  \\
\multicolumn{1}{c}{}    &                   & \multicolumn{2}{c}{CT/CT}       & \multicolumn{2}{c}{CT/CBCT}     & \multicolumn{2}{c}{T1w/T1w}     & \multicolumn{2}{c}{T1w/T1w}     & \multicolumn{2}{c}{CT/CBCT}                 \\ \cmidrule{3-12}
\multicolumn{1}{c}{}    &                   & mTRE & $\%|J|_{<0}$ & mTRE & $\%|J|_{<0}$ & DICE(\%) & $\%|J|_{<0}$ & DICE(\%) &$\%|J|_{<0}$ & DICE(\%) & $\%|J|_{<0}$            \\ 
\hline
                                                       & SyN                                        & 3.04                               & 9.8-e1                               & 57.4                                  & 0                                       & 64.5                               & 1.0e-4                              & 75.6                                & 1.5e-2                               & 78.2                                        & 0                                             \\
                                                       & uniGradICON                                & 2.07                               & 4.7e-4                               & 57.0                                  & 4.7e-4                                  & 70.6                               & 7.4e-3                              & 79.0                                & 8.9e-4                               & 81.1                                        & 6.9e-2                                        \\
                                                       & uniGradICON-$\text{LNCC}^{2}$              & 2.00                               & 0                                    & 61.0                                  & 2.8e-3                                  & 69.7                               & 2.1e-3                              & 79.6                                & 2.8e-3                               & 81.0                                        & 8.1e-2                                        \\
                                                       & multiGradICON - B                          & 2.74                               & 0                                    & 58.1                                  & 3.6e-3                                  & 69.9                               & 2.2e-4                              & 78.6                                & 1.7e-3                               & 80.9                                        & 4.1e-2                                        \\
                                                       & \cellcolor[HTML]{FD6864}multiGradICON - F* & \cellcolor[HTML]{FD6864}2.42       & \cellcolor[HTML]{FD6864}0            & \cellcolor[HTML]{FD6864}59.9          & \cellcolor[HTML]{FD6864}8.4e-3          & \cellcolor[HTML]{FD6864}71.6       & \cellcolor[HTML]{FD6864}2.9e-4      & \cellcolor[HTML]{FD6864}79.2        & \cellcolor[HTML]{FD6864}1.7e-3       & \cellcolor[HTML]{FD6864}80.5                & \cellcolor[HTML]{FD6864}2.3e-2                \\
                                                       & \cellcolor[HTML]{FD6864}multiGradICON - R* & \cellcolor[HTML]{FD6864}2.66       & \cellcolor[HTML]{FD6864}0            & \cellcolor[HTML]{FD6864}58.9          & \cellcolor[HTML]{FD6864}3.8e-2          & \cellcolor[HTML]{FD6864}70.7       & \cellcolor[HTML]{FD6864}3.4e-4      & \cellcolor[HTML]{FD6864}78.5        & \cellcolor[HTML]{FD6864}2.3e-3       & \cellcolor[HTML]{FD6864}80.6                & \cellcolor[HTML]{FD6864}5.6e-2                \\
\multirow{-7}{*}{w/o IO}                               & \cellcolor[HTML]{FD6864}multiGradICON*     & \cellcolor[HTML]{FD6864}2.27       & \cellcolor[HTML]{FD6864}0            & \cellcolor[HTML]{FD6864}58.7          & \cellcolor[HTML]{FD6864}3.2e-3          & \cellcolor[HTML]{FD6864}71.0       & \cellcolor[HTML]{FD6864}1.8e-3      & \cellcolor[HTML]{FD6864}78.7        & \cellcolor[HTML]{FD6864}2.0e-3       & \cellcolor[HTML]{FD6864}81.8                & \cellcolor[HTML]{FD6864}2.1e-2                \\ \hline
                                                       & uniGradICON                                & 1.77                               & 8.7e-5                               & 60.9                                  & 2.3e-1                                  & 70.4                               & 1.5e-3                              & 79.7                                & 6.5e-3                               & 82.2                                        & 2.4e-2                                        \\
                                                       & uniGradICON-$\text{LNCC}^{2}$              & 1.76                               & 4.8e-5                               & 62.1                                  & 2.5e-2                                  & 70.8                               & 1.6e-3                              & 80.1                                & 9.1e-3                               & 82.0                                        & 4.2e-2                                        \\
                                                       & multiGradICON - B                          & 1.84                               & 3.1e-4                               & 60.1                                  & 2.0e-2                                  & 70.8                               & 1.0e-3                              & 79.5                                & 5.6e-3                               & 82.0                                        & 1.9e-2                                        \\
                                                       & \cellcolor[HTML]{FD6864}multiGradICON - F* & \cellcolor[HTML]{FD6864}1.82       & \cellcolor[HTML]{FD6864}2.1e-4       & \cellcolor[HTML]{FD6864}61.4          & \cellcolor[HTML]{FD6864}4.3e-1          & \cellcolor[HTML]{FD6864}70.7       & \cellcolor[HTML]{FD6864}1.1e-3      & \cellcolor[HTML]{FD6864}79.9        & \cellcolor[HTML]{FD6864}6.9e-3       & \cellcolor[HTML]{FD6864}82.3                & \cellcolor[HTML]{FD6864}7.9e-3                \\
                                                       & \cellcolor[HTML]{FD6864}multiGradICON - R* & \cellcolor[HTML]{FD6864}1.86       & \cellcolor[HTML]{FD6864}2.3e-4       & \cellcolor[HTML]{FD6864}60.2          & \cellcolor[HTML]{FD6864}2.7e-1          & \cellcolor[HTML]{FD6864}70.7       & \cellcolor[HTML]{FD6864}1.4e-3      & \cellcolor[HTML]{FD6864}79.6        & \cellcolor[HTML]{FD6864}7.1e-3       & \cellcolor[HTML]{FD6864}82.1                & \cellcolor[HTML]{FD6864}8.8e-3                \\
\multirow{-6}{*}{1-$\text{LNCC}^{2}$} & \cellcolor[HTML]{FD6864}multiGradICON*     & \cellcolor[HTML]{FD6864}1.79       & \cellcolor[HTML]{FD6864}2.7e-5       & \cellcolor[HTML]{FD6864}60.6          & \cellcolor[HTML]{FD6864}2.8e-1          & \cellcolor[HTML]{FD6864}71.0       & \cellcolor[HTML]{FD6864}1.8e-3      & \cellcolor[HTML]{FD6864}79.6        & \cellcolor[HTML]{FD6864}6.3e-3       & \cellcolor[HTML]{FD6864}82.2                & \cellcolor[HTML]{FD6864}8.3e-3                \\ \hline
                                                       & uniGradICON                                & 1.87                               & 0                                    & 57.9                                  & 2.3e-2                                  & 71.7                               & 1.6e-6                              & 78.9                                & 3.4e-5                               & 82.0                                        & 1.1e-6                                        \\
                                                       & uniGradICON-$\text{LNCC}^{2}$              & 1.84                               & 0                                    & 58.3                                  & 1.8e-2                                  & 72.2                               & 8.9e-7                              & 79.3                                & 0                                    & 81.8                                        & 1.5e-5                                        \\
                                                       & multiGradICON - B                          & 1.99                               & 0                                    & 59.4                                  & 9.1e-1                                  & 72.0                               & 2.5e-7                              & 78.6                                & 3.1e-6                               & 81.5                                        & 3.4e-6                                        \\
                                                       & \cellcolor[HTML]{FD6864}multiGradICON - F* & \cellcolor[HTML]{FD6864}1.95       & \cellcolor[HTML]{FD6864}0            & \cellcolor[HTML]{FD6864}63.5          & \cellcolor[HTML]{FD6864}1.4e-3          & \cellcolor[HTML]{FD6864}71.7       & \cellcolor[HTML]{FD6864}1.0e-6      & \cellcolor[HTML]{FD6864}79.1        & \cellcolor[HTML]{FD6864}0            & \cellcolor[HTML]{FD6864}81.8                & \cellcolor[HTML]{FD6864}0                     \\
                                                       & \cellcolor[HTML]{FD6864}multiGradICON - R* & \cellcolor[HTML]{FD6864}2.03       & \cellcolor[HTML]{FD6864}0            & \cellcolor[HTML]{FD6864}63.0          & \cellcolor[HTML]{FD6864}6.6e-6          & \cellcolor[HTML]{FD6864}71.6       & \cellcolor[HTML]{FD6864}3.2e-6      & \cellcolor[HTML]{FD6864}78.7        & \cellcolor[HTML]{FD6864}2.3e-6       & \cellcolor[HTML]{FD6864}81.7                & \cellcolor[HTML]{FD6864}2.3e-7                \\
\multirow{-6}{*}{MIND-SSC}                             & \cellcolor[HTML]{FD6864}multiGradICON*     & \cellcolor[HTML]{FD6864}1.94       & \cellcolor[HTML]{FD6864}0            & \cellcolor[HTML]{FD6864}63.4          & \cellcolor[HTML]{FD6864}0               & \cellcolor[HTML]{FD6864}71.8       & \cellcolor[HTML]{FD6864}4.3e-6      & \cellcolor[HTML]{FD6864}78.2        & \cellcolor[HTML]{FD6864}0            & \cellcolor[HTML]{FD6864}81.9                & \cellcolor[HTML]{FD6864}0                     \\ \hline
\end{tabular}%
}
\end{table}

\begin{table}
\centering
\caption{Previous experimental results for Table~\ref{table:seen-multi}. An asterisk (*) indicates the affected models.}
\label{table:seen-multi-old}
\resizebox{\linewidth}{!}{%
\begin{tabular}{clcccccccccccccccccc}\toprule
\multicolumn{1}{c}{}    &                   & \multicolumn{14}{c}{Brain}                                                                                                                                                                                                                          & \multicolumn{2}{c}{Abdomen}      & \multicolumn{2}{c}{Neck to Knee}     \\
\multicolumn{1}{c}{}    &                   & \multicolumn{2}{c}{\cellcolor[RGB]{205,224,238}HCP}        & \multicolumn{12}{c}{\cellcolor[RGB]{208,234,208}Brats-Reg}                                                                                                                                                                                   & \multicolumn{2}{c}{\cellcolor[RGB]{205,224,238}Abdomen MRCT} & \multicolumn{2}{c}{\cellcolor[RGB]{255,228,206}UK Biobank}       \\
\multicolumn{1}{c}{}    &                   & \multicolumn{2}{c}{T1w/T2w}     & \multicolumn{2}{c}{T1w/T2w}       & \multicolumn{2}{c}{T1w/T1ce}     & \multicolumn{2}{c}{T1w/FLAIR}    & \multicolumn{2}{c}{T2w/T1ce}      & \multicolumn{2}{c}{T2w/FLAIR}    & \multicolumn{2}{c}{T1ce/FLAIR}  & \multicolumn{2}{c}{MR/CT}        & \multicolumn{2}{c}{FDIXON/WDIXON}  \\ \cmidrule{3-20}
\multicolumn{1}{c}{}    &                   & DICE(\%) & $\%|J|_{<0}$ & mTRE  & $\%|J|_{<0}$ & mTRE & $\%|J|_{<0}$ & mTRE & $\%|J|_{<0}$ & mTRE  & $\%|J|_{<0}$ & mTRE & $\%|J|_{<0}$ & mTRE & $\%|J|_{<0}$ & DICE(\%) &$\%|J|_{<0}$  & DICE(\%) & $\%|J|_{<0}$     \\ 
\hline
                                                 & SyN                                        & 71.9                               & 0                                   & 3.74                         & 0                              & 3.62                         & 0                              & 3.91                         & 0                              & 3.74                         & 0                              & 3.96                         & 0                              & 3.88                         & 0                              & 45.0                                   & 0                                        & 33.9                                  & 0                                       \\
                                                 & uniGradICON                                & 10.0                               & 2.6e-3                              & 13.11                        & 3.7e-3                         & 4.16                         & 2.1e-3                         & 6.38                         & 2.3e-3                         & 12.30                        & 4.6e-3                         & 8.12                         & 5.0e-3                         & 6.92                         & 3.9e-3                         & 50.0                                   & 4.0e-2                                   & 17.5                                  & 2.4e-2                                  \\
                                                 & uniGradICON-$\text{LNCC}^{2}$              & 14.5                               & 8.6e-4                              & 13.10                        & 4.9e-3                         & 4.08                         & 4.2e-3                         & 6.33                         & 3.2e-3                         & 12.24                        & 6.8e-3                         & 7.95                         & 6.0e-3                         & 6.83                         & 6.3e-3                         & 49.7                                   & 2.6e-2                                   & 18.1                                  & 4.4e-2                                  \\
                                                 & multiGradICON - B                          & 69.2                               & 8.0e-4                              & 3.60                         & 1.7e-3                         & 3.84                         & 8.1e-4                         & 4.60                         & 1.2e-3                         & 3.73                         & 2.4e-3                         & 4.95                         & 2.9e-3                         & 4.95                         & 2.3e-3                         & 58.1                                   & 3.4e-2                                   & 19.7                                  & 1.2e-2                                  \\
                                                 & \cellcolor[HTML]{FD6864}multiGradICON - F* & \cellcolor[HTML]{FD6864}40.1       & \cellcolor[HTML]{FD6864}1.0e-3      & \cellcolor[HTML]{FD6864}9.10 & \cellcolor[HTML]{FD6864}2.2e-3 & \cellcolor[HTML]{FD6864}3.66 & \cellcolor[HTML]{FD6864}1.2e-3 & \cellcolor[HTML]{FD6864}5.06 & \cellcolor[HTML]{FD6864}1.4e-3 & \cellcolor[HTML]{FD6864}8.56 & \cellcolor[HTML]{FD6864}2.8e-3 & \cellcolor[HTML]{FD6864}6.50 & \cellcolor[HTML]{FD6864}3.2e-3 & \cellcolor[HTML]{FD6864}5.60 & \cellcolor[HTML]{FD6864}2.3e-3 & \cellcolor[HTML]{FD6864}58.2           & \cellcolor[HTML]{FD6864}6.0e-2           & \cellcolor[HTML]{FD6864}20.1          & \cellcolor[HTML]{FD6864}1.2e-2          \\
                                                 & \cellcolor[HTML]{FD6864}multiGradICON - R* & \cellcolor[HTML]{FD6864}68.2       & \cellcolor[HTML]{FD6864}9.0e-4      & \cellcolor[HTML]{FD6864}3.58 & \cellcolor[HTML]{FD6864}2.3e-3 & \cellcolor[HTML]{FD6864}3.78 & \cellcolor[HTML]{FD6864}1.5e-3 & \cellcolor[HTML]{FD6864}4.48 & \cellcolor[HTML]{FD6864}2.1e-3 & \cellcolor[HTML]{FD6864}3.70 & \cellcolor[HTML]{FD6864}2.9e-3 & \cellcolor[HTML]{FD6864}4.90 & \cellcolor[HTML]{FD6864}3.8e-3 & \cellcolor[HTML]{FD6864}4.76 & \cellcolor[HTML]{FD6864}3.1e-3 & \cellcolor[HTML]{FD6864}61.1           & \cellcolor[HTML]{FD6864}8.4e-2           & \cellcolor[HTML]{FD6864}19.8          & \cellcolor[HTML]{FD6864}2.9e-2          \\
\multirow{-7}{*}{w/o IO}                         & \cellcolor[HTML]{FD6864}multiGradICON*     & \cellcolor[HTML]{FD6864}69.0       & \cellcolor[HTML]{FD6864}7.9e-4      & \cellcolor[HTML]{FD6864}3.64 & \cellcolor[HTML]{FD6864}2.5e-3 & \cellcolor[HTML]{FD6864}3.83 & \cellcolor[HTML]{FD6864}1.6e-3 & \cellcolor[HTML]{FD6864}4.68 & \cellcolor[HTML]{FD6864}1.9e-3 & \cellcolor[HTML]{FD6864}3.78 & \cellcolor[HTML]{FD6864}3.4e-3 & \cellcolor[HTML]{FD6864}5.05 & \cellcolor[HTML]{FD6864}4.1e-3 & \cellcolor[HTML]{FD6864}5.04 & \cellcolor[HTML]{FD6864}3.3e-3 & \cellcolor[HTML]{FD6864}61.8           & \cellcolor[HTML]{FD6864}4.0e-2           & \cellcolor[HTML]{FD6864}19.7          & \cellcolor[HTML]{FD6864}1.3e-2          \\ \hline
                                                 & uniGradICON                                & 7.8                                & 3.3e-3                              & 12.06                        & 3.2e-3                         & 3.63                         & 1.6e-3                         & 5.13                         & 2.0e-3                         & 11.69                        & 3.8e-3                         & 6.81                         & 4.8e-3                         & 5.71                         & 3.3e-3                         & 75.1                                   & 6.9e-1                                   & 17.0                                  & 6.9e-3                                  \\
                                                 & uniGradICON-$\text{LNCC}^{2}$              & 8.3                                & 2.4e-3                              & 11.94                        & 2.7e-3                         & 3.59                         & 1.8e-3                         & 5.05                         & 1.9e-3                         & 11.48                        & 3.2e-3                         & 6.60                         & 4.5e-3                         & 5.63                         & 3.2e-3                         & 80.3                                   & 5.8e-1                                   & 17.4                                  & 8.6e-3                                  \\
                                                 & multiGradICON - B                          & 72.8                               & 6.3e-4                              & 3.27                         & 7.4e-4                         & 3.56                         & 1.3e-3                         & 4.28                         & 1.1e-3                         & 3.34                         & 1.1e-3                         & 4.58                         & 2.5e-3                         & 4.76                         & 2.1e-3                         & 73.3                                   & 4.4e-1                                   & 18.7                                  & 5.5e-3                                  \\
                                                 & \cellcolor[HTML]{FD6864}multiGradICON - F* & \cellcolor[HTML]{FD6864}71.9       & \cellcolor[HTML]{FD6864}4.9e-4      & \cellcolor[HTML]{FD6864}3.34 & \cellcolor[HTML]{FD6864}9.5e-4 & \cellcolor[HTML]{FD6864}3.55 & \cellcolor[HTML]{FD6864}1.5e-3 & \cellcolor[HTML]{FD6864}4.29 & \cellcolor[HTML]{FD6864}1.4e-3 & \cellcolor[HTML]{FD6864}3.39 & \cellcolor[HTML]{FD6864}1.5e-3 & \cellcolor[HTML]{FD6864}4.57 & \cellcolor[HTML]{FD6864}3.1e-3 & \cellcolor[HTML]{FD6864}4.78 & \cellcolor[HTML]{FD6864}2.6e-3 & \cellcolor[HTML]{FD6864}74.8           & \cellcolor[HTML]{FD6864}3.0e-1           & \cellcolor[HTML]{FD6864}18.9          & \cellcolor[HTML]{FD6864}6.9e-3          \\
                                                 & \cellcolor[HTML]{FD6864}multiGradICON - R* & \cellcolor[HTML]{FD6864}72.8       & \cellcolor[HTML]{FD6864}5.7e-4      & \cellcolor[HTML]{FD6864}3.30 & \cellcolor[HTML]{FD6864}9.4e-4 & \cellcolor[HTML]{FD6864}3.55 & \cellcolor[HTML]{FD6864}1.5e-3 & \cellcolor[HTML]{FD6864}4.24 & \cellcolor[HTML]{FD6864}1.4e-3 & \cellcolor[HTML]{FD6864}3.37 & \cellcolor[HTML]{FD6864}1.4e-3 & \cellcolor[HTML]{FD6864}4.55 & \cellcolor[HTML]{FD6864}2.8e-3 & \cellcolor[HTML]{FD6864}4.70 & \cellcolor[HTML]{FD6864}2.5e-3 & \cellcolor[HTML]{FD6864}74.8           & \cellcolor[HTML]{FD6864}1.6e-1           & \cellcolor[HTML]{FD6864}18.6          & \cellcolor[HTML]{FD6864}6.6e-3          \\
\multirow{-6}{*}{$1-\text{LNCC}^2$} & \cellcolor[HTML]{FD6864}multiGradICON*     & \cellcolor[HTML]{FD6864}73.1       & \cellcolor[HTML]{FD6864}9.3e-4      & \cellcolor[HTML]{FD6864}3.32 & \cellcolor[HTML]{FD6864}1.1e-3 & \cellcolor[HTML]{FD6864}3.56 & \cellcolor[HTML]{FD6864}1.8e-3 & \cellcolor[HTML]{FD6864}4.27 & \cellcolor[HTML]{FD6864}1.6e-3 & \cellcolor[HTML]{FD6864}3.38 & \cellcolor[HTML]{FD6864}1.8e-3 & \cellcolor[HTML]{FD6864}4.60 & \cellcolor[HTML]{FD6864}3.2e-3 & \cellcolor[HTML]{FD6864}4.72 & \cellcolor[HTML]{FD6864}3.0e-3 & \cellcolor[HTML]{FD6864}73.3           & \cellcolor[HTML]{FD6864}2.0e-1           & \cellcolor[HTML]{FD6864}18.6          & \cellcolor[HTML]{FD6864}8.0e-3          \\ \hline
                                                 & uniGradICON                                & 39.8                               & 1.4e-5                              & 2.66                         & 0                              & 2.61                         & 8.7e-7                         & 2.81                         & 1.9e-6                         & 2.55                         & 6.7e-8                         & 2.88                         & 6.7e-8                         & 2.75                         & 7.3e-7                         & 75.4                                   & 1.9e-4                                   & 20.2                                  & 0                                       \\
                                                 & uniGradICON-$\text{LNCC}^{2}$              & 53.0                               & 9.5e-6                              & 2.58                         & 6.7e-8                         & 2.55                         & 6.7e-8                         & 2.74                         & 2.7e-7                         & 2.47                         & 6.7e-8                         & 2.80                         & 0                              & 2.68                         & 6.7e-8                         & 77.4                                   & 1.7e-3                                   & 20.8                                  & 3.1e-7                                  \\
                                                 & multiGradICON - B                          & 74.5                               & 0                                   & 2.70                         & 0                              & 2.61                         & 8.0e-7                         & 2.83                         & 6.7e-8                         & 2.62                         & 0                              & 2.94                         & 0                              & 2.79                         & 3.3e-7                         & 68.1                                   & 5.0e-3                                   & 23.3                                  & 0                                       \\
                                                 & \cellcolor[HTML]{FD6864}multiGradICON - F* & \cellcolor[HTML]{FD6864}73.0       & \cellcolor[HTML]{FD6864}9.3e-7      & \cellcolor[HTML]{FD6864}2.72 & \cellcolor[HTML]{FD6864}0      & \cellcolor[HTML]{FD6864}2.62 & \cellcolor[HTML]{FD6864}0      & \cellcolor[HTML]{FD6864}2.87 & \cellcolor[HTML]{FD6864}6.7e-8 & \cellcolor[HTML]{FD6864}2.65 & \cellcolor[HTML]{FD6864}2.7e-7 & \cellcolor[HTML]{FD6864}2.94 & \cellcolor[HTML]{FD6864}2.0e-7 & \cellcolor[HTML]{FD6864}2.82 & \cellcolor[HTML]{FD6864}0      & \cellcolor[HTML]{FD6864}70.9           & \cellcolor[HTML]{FD6864}3.2e-4           & \cellcolor[HTML]{FD6864}25.2          & \cellcolor[HTML]{FD6864}0               \\
                                                 & \cellcolor[HTML]{FD6864}multiGradICON - R* & \cellcolor[HTML]{FD6864}74.3       & \cellcolor[HTML]{FD6864}0           & \cellcolor[HTML]{FD6864}2.73 & \cellcolor[HTML]{FD6864}0      & \cellcolor[HTML]{FD6864}2.62 & \cellcolor[HTML]{FD6864}6.7e-8 & \cellcolor[HTML]{FD6864}2.88 & \cellcolor[HTML]{FD6864}6.7e-7 & \cellcolor[HTML]{FD6864}2.66 & \cellcolor[HTML]{FD6864}0      & \cellcolor[HTML]{FD6864}2.97 & \cellcolor[HTML]{FD6864}0      & \cellcolor[HTML]{FD6864}2.83 & \cellcolor[HTML]{FD6864}6.7e-8 & \cellcolor[HTML]{FD6864}70.3           & \cellcolor[HTML]{FD6864}1.0e-3           & \cellcolor[HTML]{FD6864}23.7          & \cellcolor[HTML]{FD6864}0               \\
\multirow{-6}{*}{MIND-SSC}                       & \cellcolor[HTML]{FD6864}multiGradICON*     & \cellcolor[HTML]{FD6864}74.4       & \cellcolor[HTML]{FD6864}1.4e-6      & \cellcolor[HTML]{FD6864}2.70 & \cellcolor[HTML]{FD6864}1.3e-7 & \cellcolor[HTML]{FD6864}2.62 & \cellcolor[HTML]{FD6864}1.2e-6 & \cellcolor[HTML]{FD6864}2.85 & \cellcolor[HTML]{FD6864}1.6e-6 & \cellcolor[HTML]{FD6864}2.62 & \cellcolor[HTML]{FD6864}0      & \cellcolor[HTML]{FD6864}2.94 & \cellcolor[HTML]{FD6864}1.3e-7 & \cellcolor[HTML]{FD6864}2.80 & \cellcolor[HTML]{FD6864}4.7e-7 & \cellcolor[HTML]{FD6864}70.8           & \cellcolor[HTML]{FD6864}2.7e-4           & \cellcolor[HTML]{FD6864}23.3          & \cellcolor[HTML]{FD6864}0               \\ \hline
\end{tabular}
}
\end{table}

\end{document}